\documentclass[%
 aip,
 amsmath,amssymb,
 preprint,%
]{revtex4-1}

\usepackage{graphicx}
\usepackage{dcolumn}
\usepackage{bm}

\usepackage[utf8]{inputenc}
\usepackage[T1]{fontenc}
\usepackage{etoolbox}
\usepackage{xcolor}
\usepackage{soul}
\usepackage{chemformula}
\usepackage[normalem]{ulem}
\usepackage{array}

\newcolumntype{P}[1]{>{\centering\arraybackslash}p{#1}}

\makeatletter
\def\@email#1#2{%
 \endgroup
 \patchcmd{\titleblock@produce}
  {\frontmatter@RRAPformat}
  {\frontmatter@RRAPformat{\produce@RRAP{*#1\href{mailto:#2}{#2}}}\frontmatter@RRAPformat}
  {}{}
}%
\makeatother
\begin{document}


\title{Understanding orientational disorder in crystalline assemblies of hard convex polyhedra}
\author{Sumitava Kundu}
\author{Kaustav Chakraborty}%
\author{Avisek Das}
\email{mcsad@iacs.res.in}
\affiliation{School of Chemical Sciences, Indian Association for the Cultivation of Science, Kolkata, India, 700032}

\date{\today}

\begin{abstract}
Spontaneous self-assembly of hard convex polyhedra are known to form orientationally disordered crystalline phases, where particle orientations do not follow the same pattern as the positional arrangement of the crystal. A distinct type of orientational phase with discrete rotational mobility has been reported in hard particle systems. In this paper, we present a new analysis method for characterizing orientational phase of a crystal, which is based on algorithmic detection of unique orientations. Using this method we collected complete statistics of discrete orientations along the Monte Carlo simulation trajectories and observed that particles were equally partitioned among them, with specific values of pairwise orientational differences. These features remained constant across the pressure range and did not depend on rotational mobility. The discrete mobility was characteristic of a distinct equilibrium thermodynamic phase, qualitatively different from the freely rotating plastic phase with continuous orientations. The high pressure behavior with frozen particle orientations was part of that the same description and not a non-equilibrium arrested state. We introduced a precise notion of orientational order and demonstrated that the system was maximally disordered at the level of unit cell, even though individual particles could only take few discrete orientations. We report the existence of this phase in five polyhedral shapes and in systematically curated shape families constructed around two of them. The symmetry mismatch between the particle and the crystallographic point groups was found to be a predictive indicator for the occurrence of this phase.

\end{abstract}

\maketitle

\section{Introduction}
Crystalline assemblies of colloids \cite{Pawel1983, Dinsmore1998, Li2016, Orr2022, Boles2016} and nanoparticles (NPs) \cite{Mirkin1996, Robert1996, Whitesides2002, Maye2007,Henzie2012, Bodnarchuk2011, Boneschanscher2014,  Haixin2017, Zhou2022, Geuchies2016, Shevchenko2006, Zhang2013, Jones2010, Glotzer2007} are known to form a wide varieties of simple and complex crystal structures. Structural details of colloidal crystal and nanocrystal superlattices (NCSLs) have been extensively studied over the last few decades \cite{Pawel1983, Dinsmore1998, Li2016, Boles2016,Talapin2012, Andrey2002, Tang2005, Glotzer2007, Deng2020, Samanta2022}. One aspect of the problem, namely the inherent disorders present in the otherwise regular arrangements of building blocks, is comparatively less explored \cite{Bodnarchuk2011}.  Disorders in crystalline materials have attracted significant attention in contemporary attempts to understand and engineer new properties in terms of structure \cite{Keen2015}. Therefore, it is worthwhile to explore the forms of disorders in colloidal crystalline matters or NCSLs, which may lead to the possibility of engineering desired properties as envisioned in atomic and molecular materials. 

A potential source of disorder in crystals with anisotropic building blocks is the behavior of particle orientations. It is possible to have building blocks oriented differently in the three dimensional space, while maintaining regularity in positions dictated by the crystal structure. This phenomenon is well documented in the field of molecular crystals where such phases are referred to as the rotator or plastic phase \cite{Timmermans1961, Mccullough1961, Reynolds1975, Harfenist1996, Vdovichenko2015, Even2016,Beake2017, Loidl1990, Nitta1959}, and NCSLs \cite{Dullens2007, Meijer2017, Deng2020,Elbert2021,Abbas2022,Vdovichenko2015, Even2016,Beake2017}.  Direct experimental observations of particle orientations in colloidal crystals \cite{Dullens2007, Meijer2017} and nanocrystal superlattices \cite{Deng2020,Elbert2021,Abbas2022,Hu2023} have been reported very recently, opening up the possibilities of detailed investigations of orientational disorders in the colloidal crystalline matter.

Existence of orientational disorders in crystals of colloids and nanoparticles (NPs) was predicted a decade ago by several investigators from computer simulations of idealized models \cite{Frenkel1999, John2008, Torquato2009, Miller2010, Batten2010, Ni2012, Agarwal2011a, Henzie2012,Damasceno2012c, Gantapara2013, Gantapara2015a}. In these studies, the particles were represented by convex polyhedra interacting via purely hard-core interactions, as a result, properties of the systems were solely controlled by entropy. Since then, orientationally disordered behavior have been reproduced in simulations with more realistic models \cite{Knorowski2014,Fan2019,OlveraDeLaCruz2016}. In the context of purely entropy driven crystallizations, the rotator phase has been further investigated, which led to quantitative understanding of the many-body orientational behavior in both two and three dimensions \cite{Ni2012, Gantapara2015a, Ondry2022, Shen2019, Karas2019, Lee2023}. The plastic phase in hard polyhedral systems could be termed as the ``freely rotating plastic phase'' \cite{Batten2010, Edington1999, Liu2014, Meijer2017, Harada2021, Karas2019, Damasceno2012c, Lee2023}. This meant, over an equilibrated simulation trajectory, each particle in the crystal could take all possible values of absolute orientations, while sitting at a fixed position determined by the underlying crystal structure. The rotational motions were maximally random and, in an ensemble averaged sense there was no correlations in the orientational space between a pair of particles. In spatial domain, there were only short-ranged orientational correlations, which died down beyond the immediate neighbors \cite{Agarwal2011a}. Besides that, anisotropic orientational distributions were also reported in the systems of hard superballs \cite{Torquato2009, Ni2012} and hard truncated cubes \cite{Gantapara2015a, Sharma2024}. In these studies, the particles were found to be oriented in a few preferred directions in the orientationally disordered phase.

The issue of orientational disorder in crystalline assemblies of hard convex shapes has been revisited in recent times by computer simulations, which has led to surprising discoveries \cite{Shen2019, Karas2019, Lee2023}. In a exhaustive computer simulation study on the phase behavior of hard convex polygons, Shen \textsl{et al.}\,\cite{Shen2019} has demonstrated the existence of a different type of orientational phase qualitatively different from the freely rotating plastic crystal phase mentioned above. In this phase, the particle orientations were discrete, and number of such specific orientations were fixed for a particular shape. This form of disorder was termed as the ``discrete plastic crystal'' phase by the authors, and in a subsequent study similar discrete mobility was shown in colloidal clathrate \cite{Lee2023}. In another study, Karas \textsl{et\,al.\,}\,\cite{Karas2019} reported the phase behavior of hard ``pseudo-rhombicuboctahedron (PRC)'', where six discrete orientations were observed showing distinct orientational behavior different from the freely rotating plastic phase, in that sense, exhibited clear sign of orientational correlations. The authors characterized this behavior as an ``orientational glass'' and implied that the essential differences from known phases were dynamical, and needed to be understood in the context of non-equilibrium behavior. Glassy dynamical behavior, along side discrete absolute orientations, have been long identified in certain molecular crystals, for example mixed crystals of \ch{KCN} and \ch{KBr} \cite{Berret1990, Bounds1982, Loidl1989} which were characterized as the ``orientational glasses''. Previous reports also identified the lack of continuous rotation of $\text{\ch{C}}_{60}$ fullerene molecules in the corresponding crystal structure \cite{Chaplot2001}. While hard polygons have been studied quite extensively, there is a scope for detailed investigations of the discrete rotator phase in hard polyhedra to understand the true nature of this phase, and to answer the question, if this is a common phenomenon or some rare occurrence. In real systems, multiple orientations are generally associated with glassy behavior, it is worthwhile investigating whether that is true in model systems of hard polyhedra, or the discrete plasticity in hard systems is  actually characteristics of equilibrium thermodynamic phases.

In this paper, we report both methodological advances in analysis of many-body behavior of particle orientations, as well as an extensive Monte Carlo simulations on multiple hard polyhedral systems in search of the discrete rotator phase. Our new analysis method allowed us algorithmic detection of unique orientations and it was possible to collect complete statistics of orientational fluctuations along the MC trajectories. We identified unifying features of the phase and those were used for automatic identification of orientational phase behavior. We introduced a rigorous notion of orientational order and disorder which helped us to pinpoint the global disorder in the particle orientation, even though each particle could adopt a small number of discrete orientations. Three shapes, namely Truncated Cuboctahedron (TC), Elongated Pentagonal Dipyramid (EPD) and Elongated Square Gyrobicupola (ESG) (see Fig.\,\ref{fig:pairwise_distbn_unique_orientations_snapshots}, for pictures of particle shapes) were studied in detail. The TC and ESG (also known as pseudo-rhombicuboctahedron, PRC) were chosen for benchmarking purposes in the context of published results. While identifying the phases, proper considerations for both translational and rotational motions of the particles were exercised in the context of our precisely defined notion of order and disorder in terms of particle orientations in crystalline assemblies. Our extensive search in a large set of regular convex polyhedra led to the discovery of the clear signature of discrete orientations in three more polyhedra. The robustness of the phase in a family of modified shapes around EPD and ESG as well as an empirical relationship with the certain attribute of the shape have also been reported in this work.

\section{Improved protocol for detection and analysis of orientational phase}
In this paper, we introduce an improved methodology to analyze the orientational phase behavior of a crystalline system composed of rigid bodies. The overall protocol consists of several steps, outlined below. Few of them are either already reported or quite straightforward, despite that, we include them for a coherent discussion, with proper references to previous work at appropriate places. The subsections on algorithmic detection of unique orientations and analysis of disorder at the unit cell level, are our new contributions to the overall method, and, to the best of our knowledge, have not been reported before.

\subsection{Pairwise orientational difference of two particles}\label{orien_diff}
Particle orientations were described by quaternions \cite{Anderson2016a}. The orientations of $i$-th and $j$-th particles were represented by $\mathcal{Q}_i$ and $\mathcal{Q}_j$, respectively. To estimate the orientational difference between two particles, proper consideration for the point group symmetry of the anisotropic body was considered. The point group of the convex polyhedron consisting of the rotational symmetry operations can be represented by a set of quaternions $\mathcal{Q}^p_{\gamma}$, where $\gamma = 1,2,\ldots,n_p$ and $n_p$ is the number of rotational operations in the corresponding point group. The scalar angle $\theta^p_{\gamma}$ for each rotational symmetry of the polyhedron is defined  as follows.
\begin{eqnarray}
\theta^{p}_{\gamma} = 2 \cos^{-1}[\Re(\mathcal{Q}_i^{\dagger} \mathcal{Q}_j \mathcal{Q}^p_{\gamma})] \:\:\textrm{ for }\gamma = 1,2,\ldots,n_p
\end{eqnarray}
Here the argument of the cosine inverse function means the real part of the product of three quaternions. The orientational difference between two particles denoted as $\theta_{ij}$, was defined as
\begin{eqnarray}
\theta_{ij} = \min \{\theta^{p}_1, \theta^{p}_2,\ldots, \theta^{p}_{n_P}\}
\end{eqnarray}

The quaternion angle $\theta_{ij}$ represented the minimum angle to rotate $i$-th particle into an orientation identical to that of $j$-th particle considering the corresponding point group. This definition of the orientational difference was quite standard while representing the orientation as quaternion, which was also reported in the literature \cite{Karas2019}. The histogram of the pairwise orientational differences $\theta_{ij}$ is a measure of the orientation-orientation coupling in the system, which was computed using \textsl{freud}-toolkit \cite{freud2020}. If any specific peak occurred at $\sim$ $0^{\circ}$ in the histogram profile, then the peak corresponded to the orientationally ordered particles within the certain angular tolerance. We computed the pairwise angles of all particles in the system and compared the histogram profile with complete randomness of the particle orientations in the following way. A synthetic distribution of $\theta_{ij}$ for a particular shape (point group symmetry of the corresponding particle) was measured by numerically sampling a large number of random orientations in the quaternion space using the \textsl{rowan} package \cite{rowan}, which resulted in a ``shark-fin'' shaped profile. Any specific distribution can be compared with the random profile to distinguish the difference in orientational behavior of the system. Other methods of analyzing many orientations of symmetric objects that did not use the quaternion representation have been reported in the literature \cite{Akbari2015, Nissinen2016}. Tensorial quantities called strong orientational coordinates (SOCs), introduced by Haji-Akbari \textit{et al.} \cite{Akbari2015} are particularly relevant in the current context. a brief discussion of these methods and our approach is presented in the Discussion section.

The distributions of the pairwise angles $\theta_{ij}$ were also measured at different distances to observe the spatial variation of the orientational differences between the pairs. Although, the nature of the histogram informed on the overall orientational behavior of the system, this information was not sufficient to identify the unique orientations in the system, if they were indeed present. In order to detect the unique orientations in the system, an algorithmic approach was needed, which would replace limited visual inspection generally employed in such situations. The method would require minimal intervention in the form of specification of a angular tolerance ($\theta_{c}$) to deal with the statistical noise.

\subsection{Algorithmic detection of unique orientations}\label{unique_orien}
We introduce an algorithmic way to detect unique orientations in the system, which can be applied to each frame of the simulation trajectory in a straightforward manner. The method has only one extra input parameter, namely a angular tolerance $\theta_{c}$ described here. The basic idea stems from the simple observation that two particles with the exact same absolute orientations would result in $\theta_{ij} = 0$. In practice, two particles $i,j$ were said to have the same orientations if the orientational difference was less than the tolerance, i.e.\,$\theta_{ij}\le\theta_{c}$. The value of $\theta_{c}$ depended on the shape and was well below the maximum allowed angle stipulated by the point group symmetry of the particle. In general, the first minima of the histogram of pairwise angles was considered as the value of $\theta_{c}$, which gave enough justification to consider those particles as orientationally ordered. This criterion was used to partition the particles into disjoint sets $\mathbb{O}_1, \mathbb{O}_2,\ldots,$ such that two particles belonging to the same set had $\theta_{ij}\le\theta_{c}$, and two different sets had $\theta_{ij}>\theta_{c}$. This procedure was repeated for all frames in a simulation trajectory and new sets were introduced if new unique orientations were found. Total number of unique orientations, across the entire trajectory, was the final number of the disjoint sets, $N_{\Omega}$. In this way, the identities of all particles having same orientations i.e., the pairwise angles $\theta_{ij}\le\theta_{c}$ could be monitored in any specific simulation frame. A quaternion from each disjoint set $\mathbb{O}_k$, where $k = 1,2,\ldots,N_{\Omega}$, was selected as a preferred orientation $\mathbb{Q}_k$ in the system. If a set consisted of very less number of particles, then the corresponding set could be discarded to deal with the noise.

There is an equivalent way to obtain the same information, which was amenable to better pictorial presentation of the entire data set. Here, three reference quaternions, $\mathcal{Q}_{ref, 1}$, $\mathcal{Q}_{ref, 2}$, $\mathcal{Q}_{ref, 3}$, were chosen from the system; these references were kept fixed for the entire trajectory. For the $i$-th particle, three orientational differences, $\Theta_{1, i}, \Theta_{2, i}, \Theta_{3, i}$, from the reference orientations were computed by the same formula used for the calculation of $\theta_{ij}$. In the three dimensional space spanned by the orientational differences from the fixed references, an orientation showed up as a unique point. Particles with same orientations within the tolerance $\theta_{c}$ formed clouds of points, and the number of such clouds gave the number of distinct orientations present in the system. In a noisy system, the detection of the clouds needed another clustering analysis or human intervention as an extra step. Points within a cloud had $\theta_{ij}$ values less than the cutoff, $\theta_{c}$, which explicitly depended on the polyhedral shape. This approach also served the purpose to track the identities of the particles with similar orientation, such particles belonged a single cloud. The second approach was easier for interactive analysis due to the visual representation but the first one was more quantitative and was more amenable to the automatic investigation of large data sets.

These algorithmic approaches provided the control to detect the occurrence of multiple unique orientations in the entire system as well as to estimate the population density per unique orientation with proper control of statistical noise. These methods were independent of system size which allowed us to characterize the unique orientations in a systematic way and infer important conclusions. The same analysis was equally informative for situations with continuous orientations. In this context, the three dimensional plot produced a diffused cloud of points without any sort of clustering.

\subsection{Notion of orientational order/disorder in a crystal}\label{order_disorder}
Upon successful identification of all distinct orientations present in the entire crystal, the next order of business is to properly categorize the system as either orientationally ordered or disordered. It turns out that if particle orientations are visualized in the context of underlying crystal translational symmetry, a straightforward notion of orientational order/disorder naturally emerges.

A crystal can be generated by performing translational operations on its unit cell. For a system of anisotropic particles, if the specifications of particle arrangements in a single unit cell, upon crystal translations, can reproduce the orientations of all particles in the bulk crystal, then the system is said to be orientationally ordered. If all the particles connected by the translational symmetry are not orientationally ordered with the reference particle, then it is impossible to realize the entire crystal with particle orientations obtained from a single unit cell. Such systems appears to be orientationally disordered crystal. It automatically follows from this definition that, for an ordered perfect crystal of anisotropic particles, all unit cells are translationally and orientationally identical. This precise notion of order and disorder works in the ideal crystal only. For practical implementation of this idea, one needs to define a measure of orientational attribute of a unit cell. We classified the unit cells based on the particle orientations and statistics was collected to decipher any specific pattern. One such measure is described below.

\subsection{Estimation of disorder in terms of orientational characteristics of crystal unit cells} \label{uc_stat}
After identification of the unit cells in the crystallite formed in the simulation, the unit cells were given an unique index. Following the previous step, the unique orientations of all particles were determined, i.e.\,the disjoint set the particle belonged to, at any frame of the simulation. From this data, each particle was assigned an unique index indicating the unique orientation. This index took values in the range $[0,N_{\Omega}]$. For each unit cell we associated an array of length $N_{\Omega}$, denoted by $\underline{\Lambda}$. The $w$'th element of the array was the sum of all particles in the unit cell with orientational index $w$. 
\begin{eqnarray}
\underline{\Lambda}(w) = \sum_{i\in \mathbb{U}}\delta_{o_iw}
\end{eqnarray}
where, $\mathbb{U}$ is the set of indices of particles that were part of the unit cell and $o_i$ was the orientational index of the $i$-th particle. Some elements of the array, $\underline{\Lambda}$, could be zero, since not all special orientations were present in all unit cells. Two unit cells $a$ and $b$ were said to be \textsl{orientationally identical} if and only if the following condition was true.
\begin{eqnarray}
\underline{\Lambda}^a(w) = \underline{\Lambda}^b(w) \,\,\,\textrm{for all }w.
\end{eqnarray}
Unique type of unit cells were counted and normalized distribution was calculated. The distribution of the unit cell types provided the information of orientational ordered or disordered crystals based on the population of any specific type of unit cell. In these analyses, if one kind of unit cell based on any specific arrangement of particle orientations was observed with major contribution in the system, barring the statistical noise, the system was considered to be an orientationally ordered crystal. On the other hand, if no particular unit cell type had appreciable population, the system was categorized as orientationally disordered crystal. Following this robust definition, the unit cells were categorized which helped us to determine the orientational phase in terms of broad categorization of order versus disorder, in the presence of statistical noise.

\subsection{Transition matrix for quantitative analysis of rotational mobility} \label{trans_matrix}
If the discrete hopping occurred within the unique orientations, the particles were expected to jump from one preferred orientation to another. Such type of phase was termed as ``discrete rotator'' or ``discrete plastic crystal'' as reported in the literature \cite{Shen2019, Lee2023, Chaplot2001}, qualitatively different from the continuous mobility observed in the freely rotating plastic crystal phase. In earlier investigations, the orientational hopping were characterized in terms of the discrete change of a directional vector associated with the body followed by the projection onto a unit sphere. To analyze the discrete mobility of the particles in a systematic and quantitative approach, we used the idea of transition matrix and implemented to quantify the discrete hopping. This was a straightforward tool to observe the average probabilities of discrete jump from one preferred orientation to another, over all particles in a simulation trajectory. The construction of the transition matrix in this context, is described as follows.

The transition matrix, $\underline{\underline{T}}$, is a $N_{\Omega}\times N_{\Omega}$ matrix characterizing the hopping of particles between discrete orientations along a trajectory. The number of transitions or hops between two unique orientations $\mathbb{Q}_{w}$ and $\mathbb{Q}_{v}$ for the particle $i$, $\mathcal{N}_i(w,v)$ was the number of times the particle toggled between these two orientations along the trajectory, between two consecutive frames while keeping the order preserved. The case of $w=v$ meant particle did not change orientation. The elements of particles averaged transition matrix is defined as follows.
\begin{eqnarray}
\underline{\underline{T}}(w,v) = \frac{1}{N}\sum_{i=1}^{N}\frac{\mathcal{N}_i(w,v)}{(l-1)}
\end{eqnarray}
Here, $l$ is the number of frames in the Monte Carlo trajectory. The dimension of the matrix was $N_{\Omega}\times N_{\Omega}$, where $N_{\Omega}$ was the number of unique orientations in the system. The diagonal elements would give the fraction of jumps onto the same orientations, i.e.\,an estimate of retention probability, averaged over all particles along the entire trajectory. The off-diagonal elements estimated the likelihood of hopping between distinct orientations. If discrete mobility was indeed present, one could expect to see the populations in the off-diagonal elements of the transition matrix. The data was obtained over the entire trajectory and the probability of discrete hopping from any preferred orientation to another was estimated in a more organized way, where the discrete rotator phase existed.

\section{Model and methods} \label{methods}
We study three polyhedral shapes which are known to self-assemble in crystalline solids : Truncated Cuboctahedron (TC), Elongated Pentagonal Dipyramid (EPD) and Elongated Square Gyrobicupola (ESG) \cite{Damasceno2012c,Karas2019}.  The shapes of the particles are depicted in Fig.\,\ref{fig:pairwise_distbn_unique_orientations_snapshots}A,B,C as the insets of the simulation snapshots. Considering the geometry and symmetry, the three shapes were chosen in such a way, they differed from each other sufficiently. The TC shape has 48 vertices and 26 different types of faces including the squares, hexagons and octagons with the $O_h$ point group. The EPD and ESG shapes appear to consist of 12 vertices, 15 faces and 24 vertices, 26 faces respectively. These two shapes have $D_{5h}$ and $D_{4d}$ point group respectively. We considered three more polyhedral systems: Elongated Pentagonal Gyrocupolarotunda (EPG), Elongated Pentagonal Orthocupolarotunda (EPO) and Parabiaugmented Dodecahedron (PD) and a systematically tuned shape family constructed from the ESG and EPD shapes to validate our observations. The results are discussed in the following subsections. Several standard techniques were used to investigate the translation and orientational order of the self-assembled crystals. Moreover, we developed multiple new methods to understand the orientational behavior of the particles in the crystalline solids in more systematic and precise way. Single component system of each shape, with $N = 4096$ particles, was simulated by constant pressure Monte Carlo using the Hard Particle Monte Carlo module in the HOOMD-Blue simulation toolkit \cite{Anderson2016a}. Reproducibility of all results were checked by multiple independent runs with the same sized systems. Additional simulations, by varying the system size, were also performed for $N = 2197$, $27000$ particles.

\subsection{Simulation protocol}\label{sim_protocol}
A dilute system (packing fraction $\phi$ $\sim 0.01$-$0.10$) was slowly compressed by constant volume Monte Carlo (MC) simulations up to a packing fraction of $\sim 55\%-60\%$ (the exact number depended on the particle shape). During each compression cycle, first all overlaps were removed, followed by a short MC run with the current volume. At the end of the compression stage the systems were simulated under the constant volume condition until spontaneous crystallization was observed. The crystalline solid was further equilibrated under constant pressure calculation allowing full anisotropic fluctuations of the simulation box. The reduced pressure $p^{\ast} = \beta p v_0$, where $\beta = (k_{B} T)^{-1}$ and $v_0$ is particles volume, was calculated from the constant volume crystallization simulation by the scaled distribution function method implemented in HOOMD. The particle volume $v_0$ was set to 1.0 to compare the system density with packing fraction. This value of pressure was taken as a base value and the system was further compressed by slowly ramping the external pressure, at each pressure value long equilibration run was carried out to ensure stable fluctuations of the box dimensions around the mean values. This process was continued to the a high value of external pressure beyond which no noticeable decrease in volume was observed. The entire process typically involved $\sim 25-27$ stages. The most compressed state obtained at the highest value of pressure was taken as the starting point for a series of fresh set of constant pressure simulations  where the pressure was reduced slowly, until the system completely melted into an isotropic liquid phase. The entire melting regime was simulated at same pressure values where the compression simulations were performed. The production part of the trajectory at each packing fraction in the melting runs was used to perform all analysis.

The detailed phase behavior of the EPD and ESG systems were studied thoroughly and the data were collected from the compression and melting runs. The hysteresis analyses were performed for the two shapes and rest of the analyses were carried out using the data of melting simulations only. To investigate the translational order, in addition to the standard analyses, we detected the unit cells of the crystal obtained from the simulations. The identified unit cells were verified with the reported crystal structures of the respective polyhedral shapes and perfect agreement was observed in each system. The orientational phase behavior was analyzed by our newly introduced protocol.

\section{Results} \label{results}
The Truncated Cuboctahedron (TC) self-assembled into body-centered tetragonal (BCT) crystal structure upon compression of the face-centered cubic (FCC) plastic crystal phase. The other two shapes EPD and ESG crystallized in the FCC structures for the entire solid region. The translational and orientational behavior of the TC and ESG shapes were already reported in the entire solid region \cite{Karas2019}. Our independent investigation confirmed the existing results; the TC shape exhibited first order transition between orientationally ordered (two unique orientations) BCT structure and FCC plastic phase, where as, the ESG shape maintained six unique orientations in the FCC crystal at the densest state and transitioned into FCC plastic phase at lower packing fractions before melting to the isotropic liquid phase. Here, we used these two shapes for benchmarking and implemented the newly developed analyses protocol on all the three systems. The detail data is reported for the systems of TC, EPD and ESG shapes. Similar characterization was carried out for three additional shapes and modified shape families of EPD and ESG shapes for the validation purpose. In the light of new analyses techniques, we report the detail orientational behavior of anisotropic polyhedra in the crystalline solids.

\begin{figure*}
	\centering 
	\includegraphics[scale=0.22]{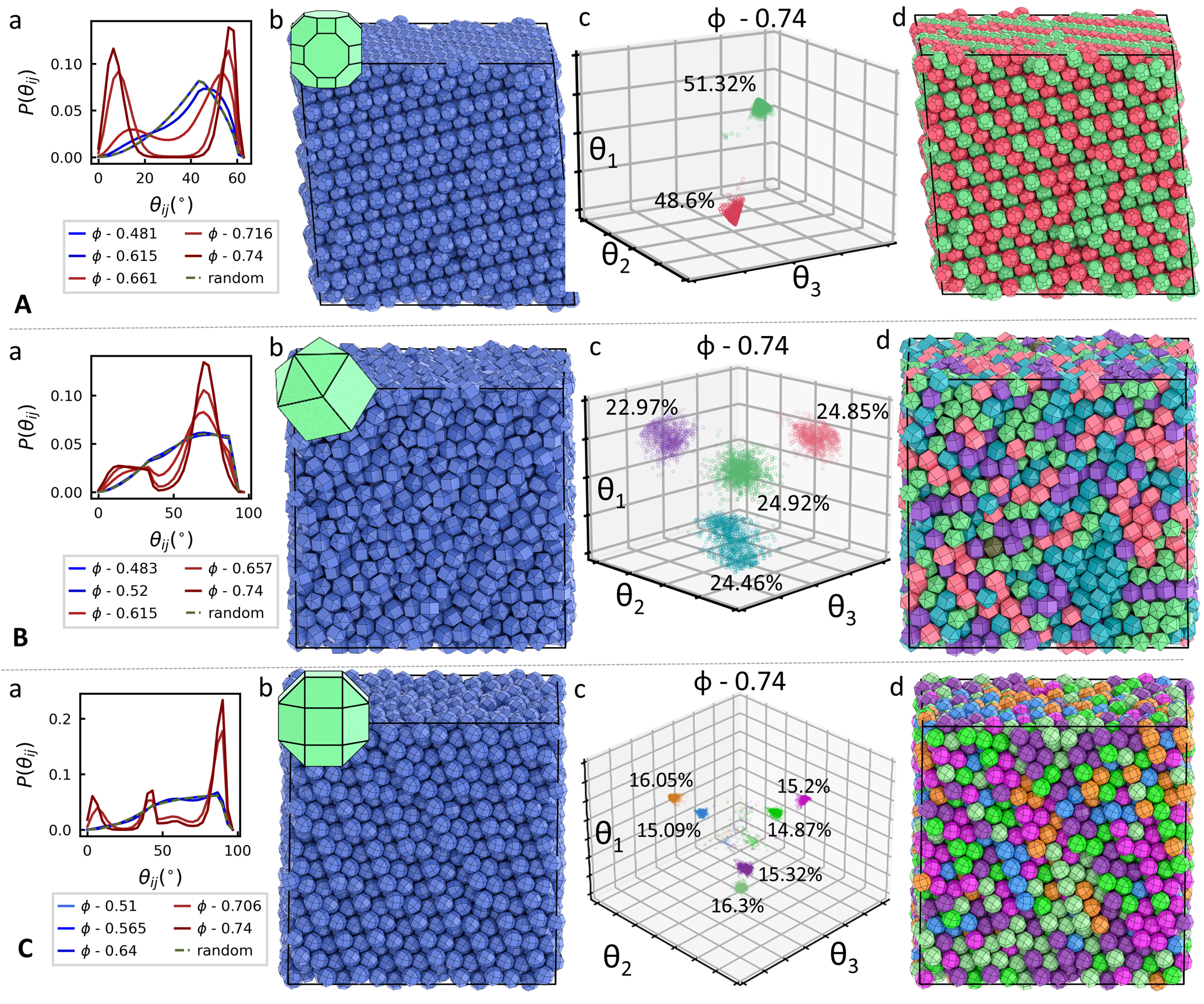}
	\caption{\textbf{Unique orientations and simulation snapshots at high packing fractions are shown for three shapes:} (\textbf{A}) Truncated Cuboctahedron (TC), (\textbf{B}) Elongated Pentagonal Dipyramid (EPD) and (\textbf{C}) Elongated Square Gyrobicupola (ESG) are shown. Results presented in this figure corresponded to roughly the same packing fraction. Panel (a) represents the distributions of all pairwise angles in the system at different packing fractions. The simulation snapshots at the densest packing ($\phi \sim\,0.74$ for all shapes) are shown in the panel (b). The panel (c) of each subfigure depicts the unique orientations of all particles in the systems in a three dimensional space of orientational differences with the population shown in percentage. Particles with the same orientations show up as distinct clusters with unique color code. Figures in panel (d) show the snapshots of same systems with uniquely oriented particles rendered in a single color. The TC system has two orientations where as EPD and ESG systems have four and six orientations respectively with equal population densities.}
	\label{fig:pairwise_distbn_unique_orientations_snapshots}
\end{figure*}

\subsection{Orientational behavior of particles and solid-solid phase transitions}
Ensemble averaged distribution of all pairwise orientational differences, $\theta_{ij}$, is a reporter of the orientational phase of the bulk crystal obtained in the MC simulations while compared to the random ``shark-fin'' profile as discussed in the Section \ref{orien_diff}. The orientational phase diagrams represented in terms of distributions of $\theta_{ij}$ are shown in the panel (a) of Figs.\,\ref{fig:pairwise_distbn_unique_orientations_snapshots}A,\,B,\,C. Each of the three shapes had a completely random orientational phase i.e., plastic crystal, characterized by the ``shark-fin'' shaped profiles, at low pressures. The plastic phase persisted within a pressure range, then there were qualitative changes in the distributions of $\theta_{ij}$ beyond the packing fractions $\sim0.65$, $\sim0.621$ and $\sim0.65$ for the TC, EPD and ESG systems, respectively (compare the graphs in each plot in the panel (a) of Fig.\,\ref{fig:pairwise_distbn_unique_orientations_snapshots}). The distributions in the high pressure regions were characterized by a small number of peaks, with widths that decreased as the external pressures were gradually increased. The TC shape had two prominent peaks with equal intensities (panel (a) of Fig.\,\ref{fig:pairwise_distbn_unique_orientations_snapshots}A). The EPD shape gave two peaks with widely different heights. The smaller one had less well-defined features (panel (a) of Fig.\,\ref{fig:pairwise_distbn_unique_orientations_snapshots}B). There were three peaks for the ESG shape; two peaks had equal population and another one existed with more population than the previous two (panel (a) of Fig.\,\ref{fig:pairwise_distbn_unique_orientations_snapshots}C). At the highest achievable pressure for each of these systems, cleanest profile, characterized by narrowest peak widths, was obtained. Multiple peaks in the orientational distributions were reported in the phase diagram of water \cite{Aragones2009}, simulations of hard dumbbells \cite{Vega1992}, hard polyhedral particles \cite{Gantapara2015a, Karas2019, Lee2023} and hard  polygons \cite{Shen2019}. The height differences of the peaks in the distribution profiles indicated the existence of the particles with multiple unique orientations corresponding to the peak observed in the extreme higher values of $\theta_{ij}$. Despite the noise, for all three shapes, the qualitative natures of the distributions were retained across the entirety of the high pressure regions. This distribution profile can be regarded as an order parameter for the many-body orientational states of the system to distinguish any different orientational behavior from the complete random phase.

These results suggested a solid-solid phase transitions involving the behavior of the particle orientations. For the TC shape, the orientational transition coincided with the structural phase transition between the FCC and BCT crystal structures \cite{Karas2019}. For the other two shapes, however, the transitions were purely orientational in nature and no changes in the positional structures of the solids were detected. This behavior for the ESG shape has already been reported \cite{Karas2019} which is also verified in our independent investigations using the newly proposed analyses protocols. The observations in the EPD system were qualitatively identical, suggesting that solid-solid transitions of purely orientational type, sharing the above mentioned qualitative features, could be a generic phenomenon and deserved further analysis.

All the orientational behavior were quantitatively reproducible, as multiple independent runs of the same system size produced virtually identical distributions. Profiles different from the typical shark-fin shapes of completely random orientations, indicated the presence of inter-particle correlations in the orientational space. The allowed differences, i.e.\,possible values of $\theta_{ij}$ where the distributions peaked were conserved in multiple runs, suggesting that they were intrinsic properties of the shape. As mentioned before, the qualitative features of the distributions remained intact in the entire high pressure region, before the appearance of the plastic crystal.

\subsection{Existence of unique orientations and their statistical distributions}
As the distribution profiles of all pairwise angles $\theta_{ij}$ did not elucidate the number of unique orientations, our newly developed algorithmic protocol was implemented on these systems as discussed in the section \ref{unique_orien}. Unique orientation detection in the TC system at high packing fraction in the BCT phase confirmed the known fact of two distinct orientations (see two clouds of points in the panel (c) of Fig.\,\ref{fig:pairwise_distbn_unique_orientations_snapshots}A). Upon coloring the two distinct values as red and green, the system was found to be composed of two colored particles arranged in a regular manner (panel (d) in Fig.\,\ref{fig:pairwise_distbn_unique_orientations_snapshots}A). But for the EPD and ESG systems, careful visual inspection was enough to distinguish the lack of order in the EPD and ESG systems, different from the scenario of orientational order in the TC system. Algorithmic detection of unique orientations on the EPD and ESG shapes revealed that both these systems were composed of few discrete orientations for the overwhelming majority of particles. The numbers of unique orientations were four and six for the EPD and ESG shapes, respectively. The existence of multiple unique orientations in the crystal structure was already reported in computer simulations \cite{Aragones2009, Vega1992, Ni2012, Shen2019, Karas2019, Lee2023, Sharma2024} and experiments of colloidal and nanoparticles \cite{Berret1990, Bounds1982, Abbas2022}. The clouds of points in the orientational differences from three reference orientations supported these conclusions (Figs.\,\ref{fig:pairwise_distbn_unique_orientations_snapshots}B,C panel (c) of each panel). The same plots in the plastic crystal phase, with the same three reference orientations, had an almost homogeneous distributions of points, confirming the random nature of orientations in the allowed space, consistent with the freely rotating picture mentioned before. By our definition, particles with the same unique orientations, i.e.\,points within a cloud, would have $\theta_{ij}$ values less than the cutoff $\theta_{c}$ and \textsl{vice versa}. The values of $\theta_{c}$ were set to $40^{\circ}$ and $20^{\circ}$ for EPD and ESG shape respectively. The values of the cutoff angle could be rationalized by the positions of the first peak near zero in the distributions of the $\theta_{ij}$. This peak corresponded to the particles with same absolute orientations. The fact that a single value of $\theta_{c}$ was sufficient for the entire system could be justified from the qualitatively similar nature of properly calculated per-particle orientational fluctuations observed in both the systems. This behavior is expected for an equilibrium crystal where each particle has identical local environment, which was the case here, as both the systems adopted an FCC structure. Overall, the data was consistent with the global distributions of $\theta_{ij}$ in the panel (a) of each panel of Figs.\,\ref{fig:pairwise_distbn_unique_orientations_snapshots}B,C. Moreover, the unique orientations in the ESG shape appeared to have a pattern of three sets of pairs (panel (c) of Fig.\,\ref{fig:pairwise_distbn_unique_orientations_snapshots}C) indicating the signature of another inter-particle orientational coupling and its consequence will be discussed later in the paper. For both systems, particles with same absolute angular dispositions are shown in same color and the arrangements of the different orientations are portrayed in bottom panels of  Fig.\,\ref{fig:pairwise_distbn_unique_orientations_snapshots}B and C. Partitioning of the particles among the distinct orientation was found to be equal, barring statistical noise, for both the systems (see panel (c) of Figs.\,\ref{fig:pairwise_distbn_unique_orientations_snapshots}B,C). All these features of this phase were reproducible in multiple independent simulations with the same sized systems.

\begin{figure*}
	\centering
	\includegraphics[scale=0.95]{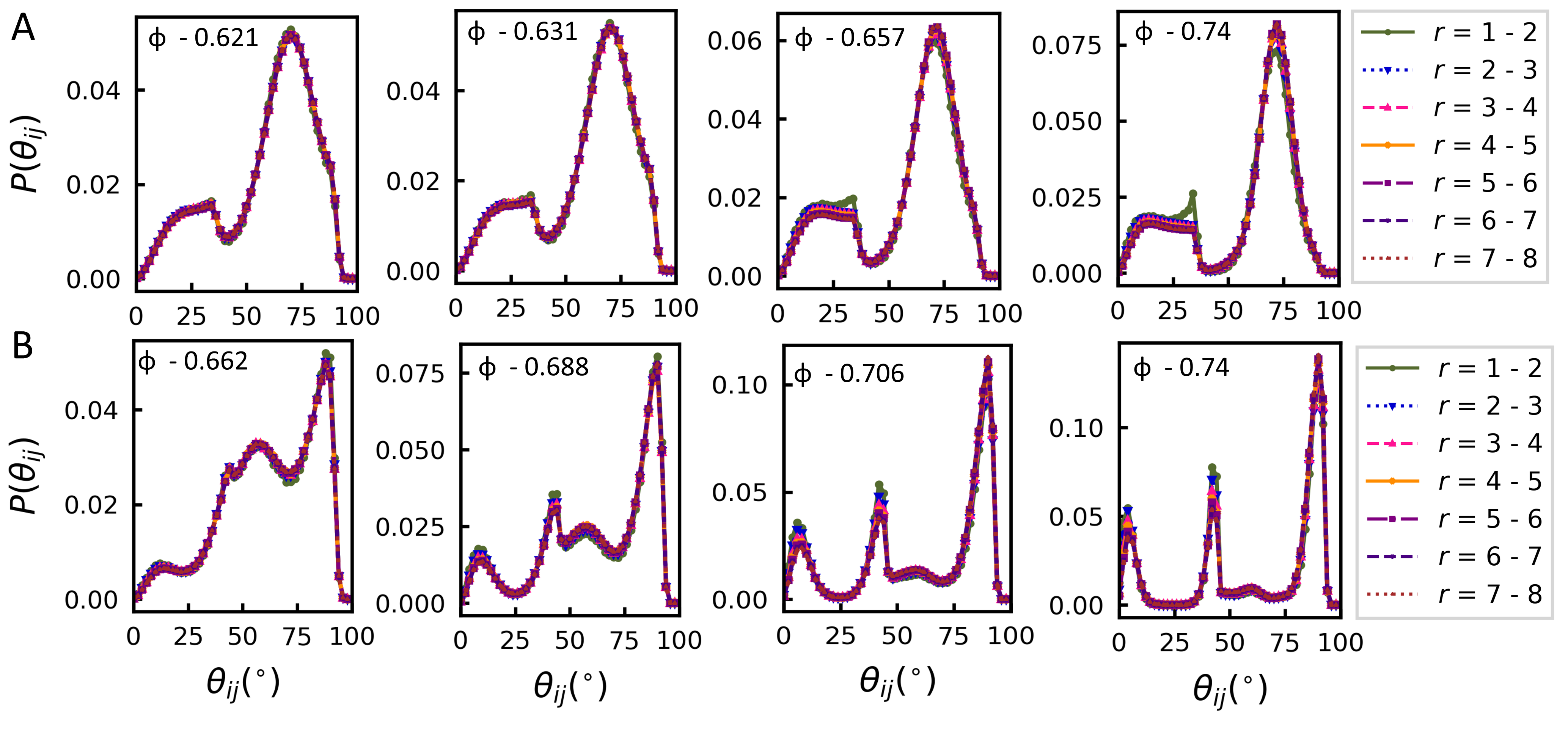}
	\caption{\textbf{Spatial dependence of the distribution of pairwise angles at four packing fractions are shown for two shapes at different pairwise distances:} \textbf{(A)} EPD and \textbf{(B)} ESG. All the distributions appear to be similar at different distances ($r$) with fixed $\Delta r$ at particular packing fraction indicating the homogeneous distribution of each unique orientations in the system.}
	\label{fig:spatial_dependence}
\end{figure*}

\subsection{Spatial distribution of unique orientations}\label{dist_inv}
To observe the spatial dependence of the unique orientations in the disordered solids of the EPD and ESG shapes, we explored the ensemble averaged distributions of pairwise angles ($\theta_{ij}$) among the particles located at the distances ($r_{ij}$) between $r$ to $(r + \Delta r)$ from a central particle. The pairwise distance $r$ (= $r_{ij}$) was varied up to the system length scale to study the distance dependent behavior of the inter-particle orientational coupling as shown in Figs.\ref{fig:spatial_dependence}A,B. The data were plotted at four packing fractions for each shape. For a particular packing fraction $\phi$, for both shapes, the distributions at various distances appeared to be virtually identical barring statistical noise. This fact suggested no significant changes in the orientational behavior at any specific distance, ruling out spatial clustering of distinct orientations. For the EPD shape shown in Fig.\ref{fig:spatial_dependence}A, all the distributions exhibited similar nature with two peaks having sufficient height difference as shown in the panel (a) of Fig.\,\ref{fig:pairwise_distbn_unique_orientations_snapshots}B. The width of the peaks broadened as the packing fraction decreased and the broadening of all peaks in the corresponding histogram at a packing fraction, maintained the distance independent behavior. Similarly, Fig.\ref{fig:spatial_dependence}B also suggests the constant nature of the distributions of pairwise angles for the ESG shape at different packing fractions producing three peaks as observed in the panel (a) of Fig.\,\ref{fig:pairwise_distbn_unique_orientations_snapshots}C. The quantitative differences between the distance invariant distributions corresponding to the chosen packing fractions could be explained by rotational mobility of the particles discussed later in the manuscript. These data provided strong indication of the existence of long-range orientational correlation in the disordered state. This pattern remained conserved up to certain packing fractions where the phase existed exhibiting the discrete rotation at the highest extent, before the system transitioned into the freely rotating plastic crystal.

In the light of this finding, we could explain the identical nature of distributions of $\theta_{ij}$ in both cases, namely within the neighboring particles as  well as the considering all pairs irrespective of inter-particle distances. In summary, all particles exhibited similar kind of local orientational environments across the bulk phase, further justifying the use of a single value of $\theta_{c}$ in the detection of unique orientations.

\begin{figure*}
	\centering
	\includegraphics[scale=0.9]{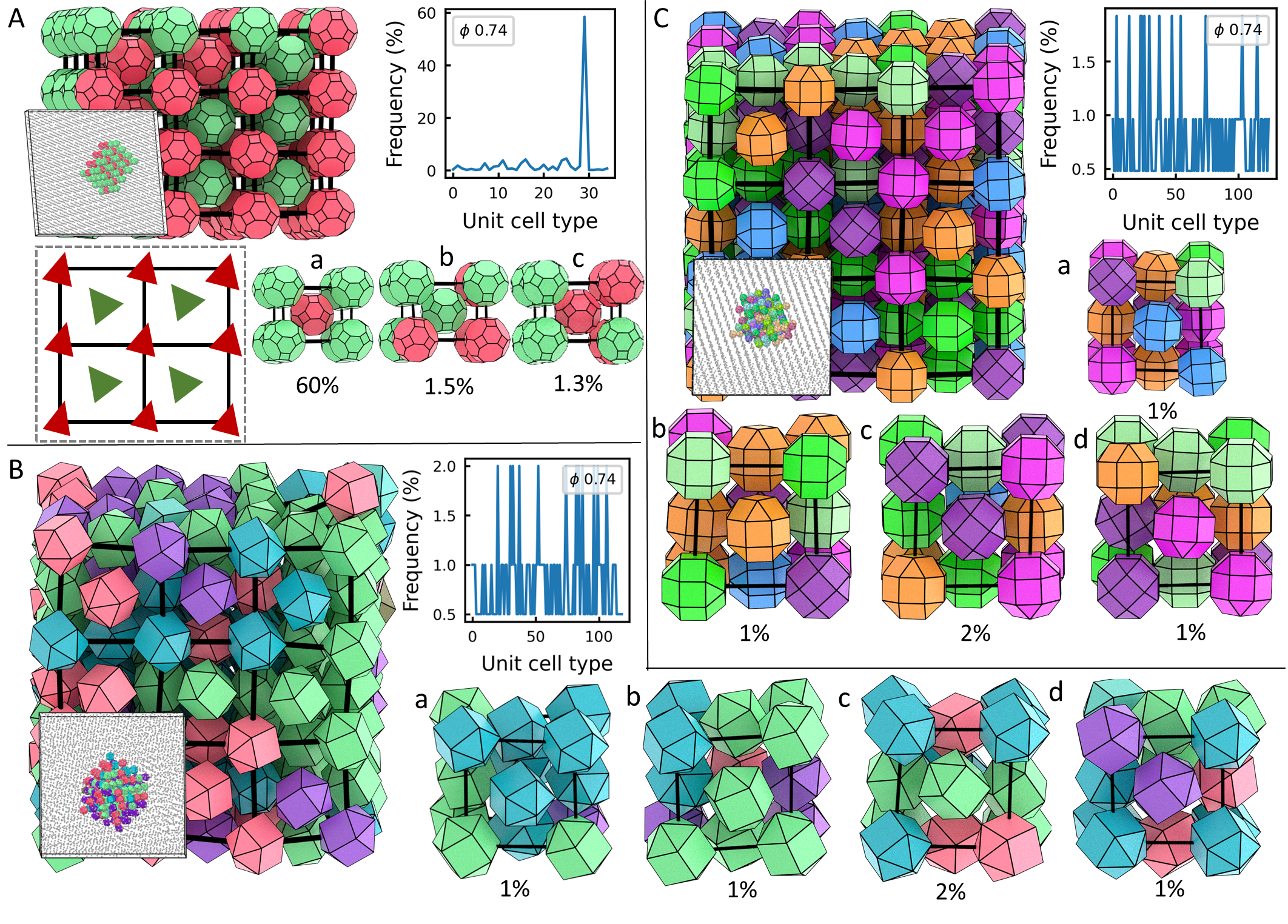}
	\caption{\textbf{Orientational order/disorder and crystal unit cell}. The notion of orientationally ordered crystal is illustrated in the cartoon at the lower left corner of panel \textbf{A}. In the 2d body centered square lattice with one particle per lattice site, particles at translationally connected positions have the same orientations. Consequently, all the unit cells are orientationally identical, with four red orientations at the corners and a green orientation at the center. Rest of the figure presents orientational analysis of unit cells at the maximum packing fractions. Data for TC, EPD and ESG shapes are shown in panels \textbf{A}, \textbf{B} and \textbf{C}, respectively. Particles are colored by their absolute orientations as in Fig.\,\ref{fig:pairwise_distbn_unique_orientations_snapshots}. Small portion of the simulation system with the unit cells outlined, is shown in each panel. The relative position of the isolated part in the whole system is indicated in the inset. All positions and orientations were exactly reproduced from the simulation data. Statistics of orientationally categorized unit cells are shown in the plots (for details, see the main text). The TC shape had only one type of unit cell, but both EPD and ESG crystals were composed of large numbers of orientationally diverse unit cells. The nature of disorder in EPD and ESG, as opposed to ordered arrangement in TC, is clearly evident from this figure.}
	\label{fig:unit_cell}
\end{figure*}

\subsection{Nature of disorder in terms of crystal unit cells} 
The nature of orientational behavior in the systems became apparent in the light of defining the notion of ordered and disordered crystal based on the distinct orientations. According to the notion mentioned in section \ref{order_disorder}, in an orientationally ordered perfect crystal of anisotropic particles, all unit cells are translationally and orientationally identical (see the cartoon in panel A of Fig.\,\ref{fig:unit_cell}). We tested this idea on the TC system by mapping the crystal into unit cells (Fig.\,\ref{fig:unit_cell}A top corner depicts a chunk of the simulation system with unit cells information shown as lines connecting the lattice points). The TC system had a unit cell with one particle at the center and eight particles at the corners of a parallelepiped. It was found that only one type of unit cell, where all the corner particles had one orientation and the center particle adopted the second orientation, dominated the system (plot in Fig.\,\ref{fig:unit_cell}A, and unit cell type (a) with 60\% abundance). Other unit cell types, for example, (b) and (c) shown in Fig.\,\ref{fig:unit_cell}A, occurred as noise. These data confirmed our notion of orientationally ordered crystal, where particle orientations were dictated by translational relationships characteristic of the crystal structure.

Unit cell analysis of EPD and ESG systems gave very different results (Fig.\,\ref{fig:unit_cell}B and C). The orientation of the EPD and ESG shapes with respect to the corresponding FCC lattice were shown at the highest packing fractions ($\phi$ $\sim$ 0.74 for both shapes) in the form of unit cells consisting of fourteen particles. It was apparent that all three particles connected via crystal translational vectors with a reference one, did not share the same absolute orientations, as found in the TC system, ignoring statistical noise. But all the particles in the unit cell did choose any one of the unique orientations found in the crystal structure. The FCC unit cells were orientationally diverse, and no single type dominated across the bulk systems (illustrative examples are shown in (a)-(e) of panels B and C in Fig.\,\ref{fig:unit_cell}). The statistics of unit cell types showed that both systems were composed of large number of orientationally distinct unit cells, consequently, population of a single type was negligibly small, often amounting to few in number (plots in Fig.\,\ref{fig:unit_cell}B and C). Not all distinct orientations were present in every unit cell. At the level of unit cells, the system looked random. This fact was consistent with the visual impressions of the orientational disorder in the snapshots of the whole systems (Fig.\,\ref{fig:pairwise_distbn_unique_orientations_snapshots}B,C). Knowing the positions and orientations of the particles in any one of the unit cells was not sufficient to reconstruct the entire crystal as realized in the simulation. In other words, particle orientations did not maintain any discernible relationship with the translational symmetry of the crystal, a situation exactly opposite to the orientationally ordered crystal in TC. Though the orientational disorder with six unique orientations was reported in the ESG system in earlier investigation by Karas \textit{et al.} \cite{Karas2019}, our approach to categorize the unit cells based on the particle orientations indicated the true nature of orientational disorder in the EPD and ESG systems. The disorder was limited at the level of single particle, which took any one of the preferred orientations. But the distinct orientations were distributed over the system in such a way, orientational attributes of a single unit cell was not enough to construct the entire crystal. Hence, the disorder was non-random at the level of single particle but random at the level of unit cells. The detection of the unit cells and binning those based on the orientational environment served the purpose of straightforward identification of the orientationally ordered or disordered phase in crystals. This observation allowed us to study the orientational phase in the crystalline structure confirming the essential differences between the ordered and disordered crystals.

\begin{figure*}
	\centering
	\includegraphics[scale=0.9]{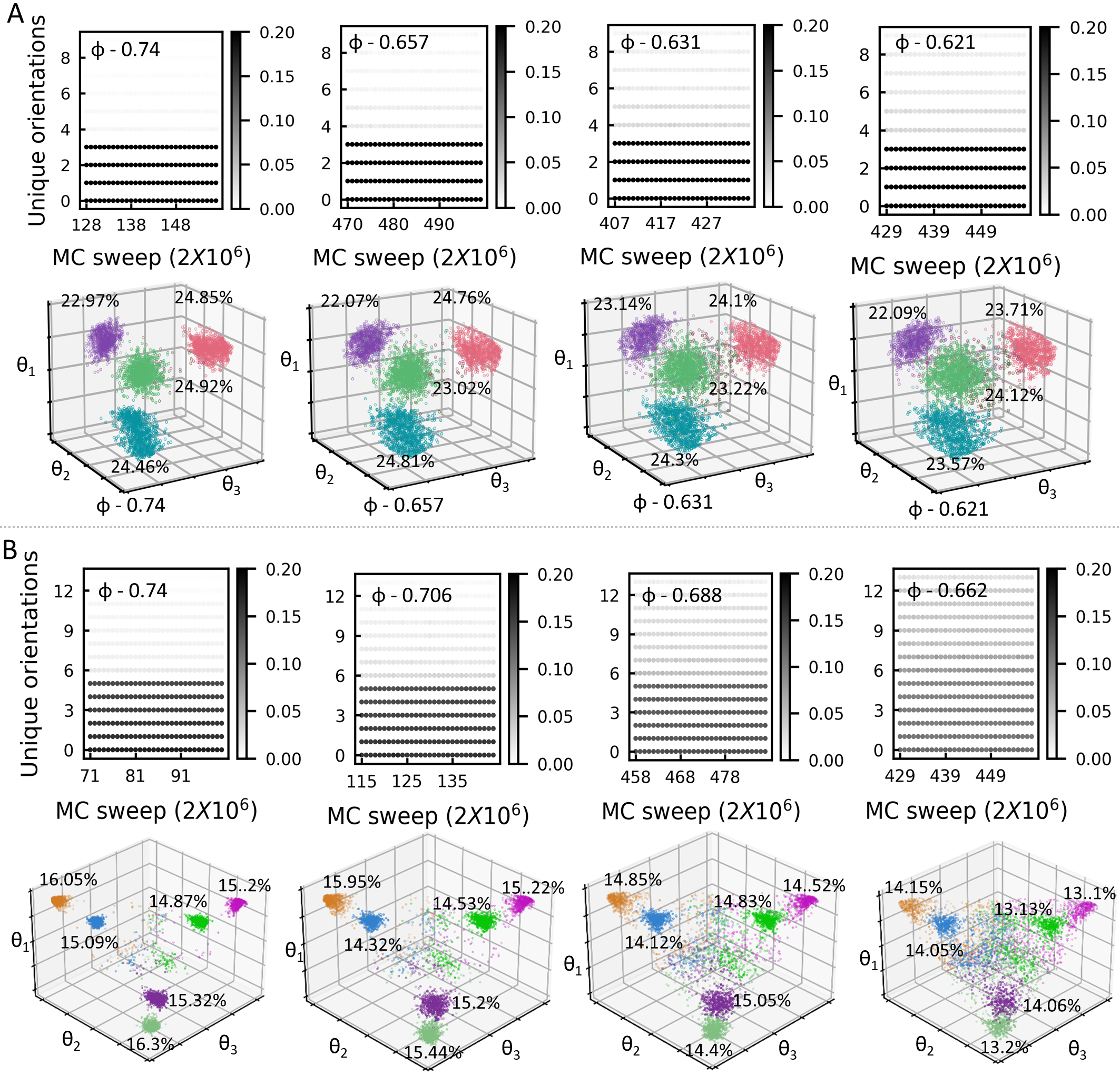}
	\caption{\textbf{Fixed number of absolute unique orientations in the system keeping the pairwise angle distribution intact over the trajectory are shown for four packing fractions:} \textbf{A.} EPD and \textbf{B.} ESG shapes. Top row of each panel depicts the fixed number of unique orientations over the trajectory at four packing fractions as melting sequence. The second row represents the data at the same packing fractions with fixed number of unique orientations keeping the positions of the clouds almost intact in the three dimensional space of $\Theta_1, \Theta_2, \Theta_3$. Each cloud has been designated with a unique color representing almost same absolute orientations under a certain angle tolerance of the particles in the system. The number of unique orientations appear to be four and six for EPD and ESG shapes respectively. Almost equal compartmentalization of the system particles is also shown with the numbers in percentage for the two shapes at the four packing fractions.}
	\label{fig:discrete_mobility_unique_orientations}
\end{figure*}

\subsection{Discrete rotational mobility of the disordered phase}
The qualitative similarities of distributions of $\theta_{ij}$'s (Figs.\,\ref{fig:pairwise_distbn_unique_orientations_snapshots}B,C, first row) suggested that the unique orientations were present all across the pressure range where this phase was present. The increased widths of the peaks could possibly mean elevated level of rotational mobility of the particles. In order to understand the pressure dependence of the orientational behavior at each packing fraction, we looked at the unique orientations calculated using the similar protocol and possible changes of orientations along the simulation trajectories.

At each packing fraction, we counted the total number of unique orientations ($N_{\Omega}$) sampled by all particles over the MC trajectory. Absolute nature of those orientations were verified by choosing the same set of three reference orientations. With these choices, points in the same regions of the three dimensional space spanned by $\Theta_1$, $\Theta_2$, and $\Theta_3$ meant the occurrence of the same number of orientations with fixed angular differences. The results for several packing fractions during the melting of the high pressure crystals are shown in Fig.\,\ref{fig:discrete_mobility_unique_orientations}. For lower density solids, $N_{\Omega}$ slowly increased (Fig.\,\ref{fig:discrete_mobility_unique_orientations}A,B top panels). The ESG system adopted slightly higher number of orientations at the lowest packing fraction before the appearance of the freely rotating plastic phase. But at each state point, the majority of the particles belonged to four and six orientations for the EPD and the ESG shapes, respectively. Locations of the clouds of points in the $\Theta_1$, $\Theta_2$, $\Theta_3$ spaces confirmed that same set of discrete orientations appeared in all values of pressure, across the stability range of this phase. Not only the same values appeared, the relative populations among them were also invariant of the packing fraction (first and second row of Figs.\,\ref{fig:discrete_mobility_unique_orientations}A,B). This data was consistent with the qualitative similarities of the profiles of distributions of $\theta_{ij}$'s (Figs.\,\ref{fig:pairwise_distbn_unique_orientations_snapshots}B,C, first row). The negligible populations of extra orientations could be disregarded as statistical noise. The entire data set strongly suggested that the number of unique orientations, equal populations, and constant orientational differences i.e., the positions of the peaks in the $\theta_{ij}$ plots, were defining features of the phase, as they remained constant.

In terms of rotational mobility of particles fixed at their respective lattice sites, two possibilities could be consistent with the data. Either the particles stayed in the same orientations with larger amplitudes of motion, or there were discrete jump motions between the selected orientations, in a correlated manner, such that the total partitioning remained constant across the simulation trajectory (Fig.\,\ref{fig:discrete_mobility_unique_orientations}). The latter possibility would mean the existence of ``discrete mobility'' qualitatively different from the continuous mobility observed in the freely rotating plastic crystal phase. However, it is important to note that both scenarios would be consistent with the qualitatively invariant nature of these distributions. The discrete orientational hopping in the crystal is not new and was already reported in the entropy driven systems of hard polygons \cite{Shen2019} and hard polyhedra \cite{Lee2023}. Such orientational phase was properly characterized in those studies and the phase was termed as ``discrete rotator'' or ``discrete plastic crystals''. 

\begin{figure*}
	\centering
	\includegraphics[scale=0.95]{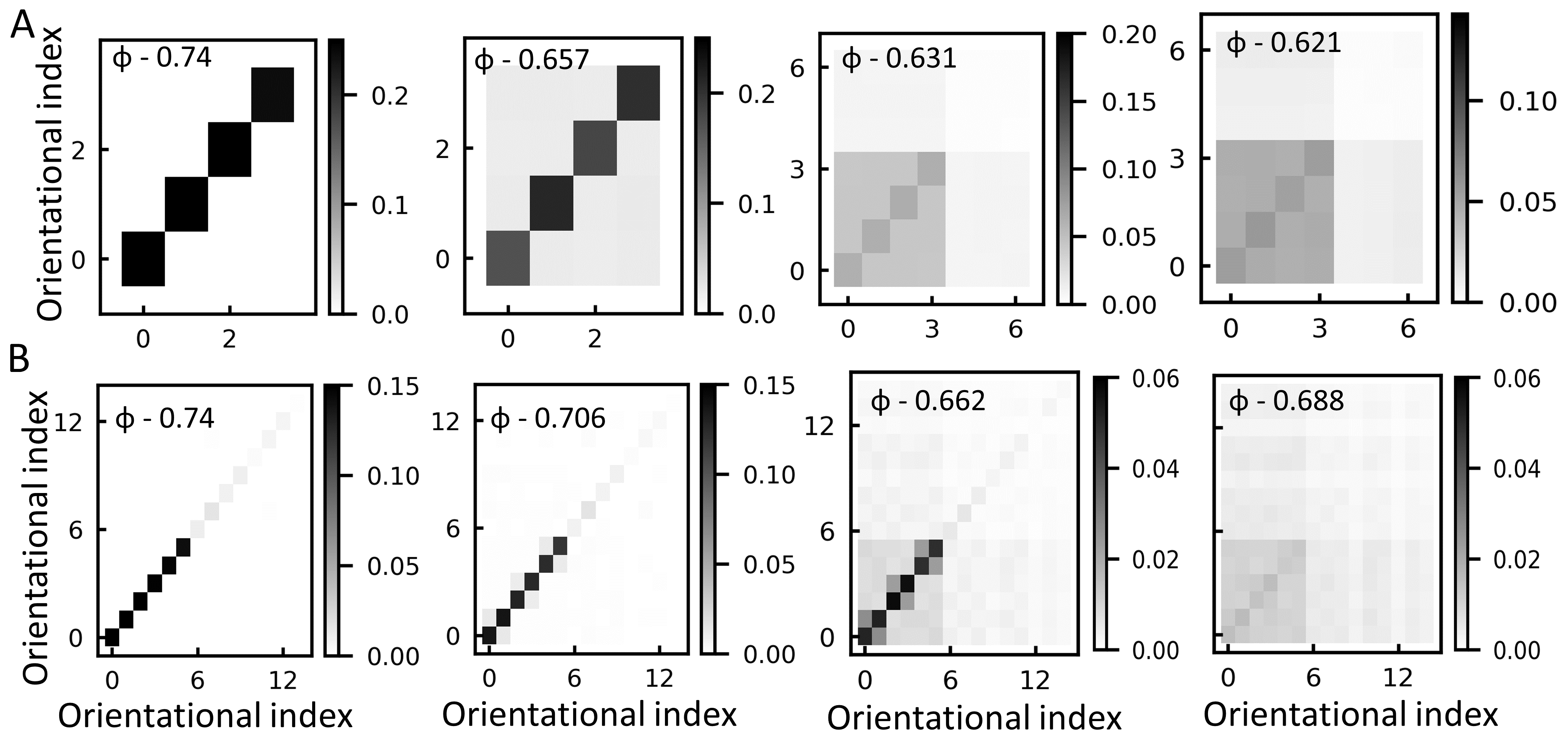}
	\caption{\textbf{Signature of orientational hopping with decreasing packing fractions are shown.} \textbf{A.} EPD and \textbf{B.} ESG shapes. The data for transition matrix is shown for the two systems upon decreasing packing fractions. The matrix represents the probability of hopping for particles from one unique orientations to another one. At the densest packing only the diagonal part of the matrices are populated describing there was no hopping at the highest packing fractions. Upon melting the particles start to toggle within the unique orientations maintaining the probability of hopping to any other orientations except the special ones is almost negligible. \textbf{(B)} depicts the ESG particles toggle within two special orientations considered as pair of clouds at $\phi$ $\sim$ 0.706. Upon melting, the probability of hopping for ESG particles increases beyond the pairs resulting the transition matrix almost picks up homogeneous distribution at $\phi$ $\sim$ 0.662 within six unique orientations. Hopping matrices in the top and bottom panel of each sub-figure \textbf{(A)} and \textbf{(B)} are shown in different ranges.}
	\label{fig:discrete_mobility_hopping_matrix}
\end{figure*}

We calculated the transition matrices to estimate the average probabilities of jumping from one orientation to another, over a Monte Carlo trajectory as discussed in the section \ref{trans_matrix}. These matrices were calculated and averaged over 500 simulation frames, which were written with a period of $2\times 10^{5}$ Monte Carlo Sweeps (MCS), along production runs with lengths of 100 million MCS. The data for both the shapes are shown in Fig.\,\ref{fig:discrete_mobility_hopping_matrix}, which validated the second scenario. In the maximum packing fractions there were no hopping (no off-diagonal population at $\phi=0.74$, Figs.\,\ref{fig:discrete_mobility_hopping_matrix}A,B), which suggested that at highest pressures, the orientations were frozen in one of the discrete values. With decreasing packing fractions, off-diagonal elements picked up more populations, even though particles still had a greater tendency to remain fixed in one of the preferred values, resulting in higher contributions along the diagonal. At sufficiently low packing fractions, all chosen orientations were sampled by the same particle with equal probability and all the off-diagonal elements obtained appreciable populations ($\phi-0.621$ and $\phi-0.662$ for the EPD and ESG shapes, respectively). This trend continued, and just before the melting into fully rotating plastic phase, maximum hopping was observed. Transitions between a preferred and a new orientation was always much less frequent and could be treated as noise all over the phase diagram where this phase was identified (Figs.\,\ref{fig:discrete_mobility_hopping_matrix}A,B). These quantitative observations produced by the new analysis protocol agreed with the characteristics of the discrete rotator phase as reported earlier \cite{Shen2019, Lee2023}. The data provided direct proof of discrete mobility of the disordered state along with the hopping probabilities within the preferred orientations. 

At low density solids in the discrete rotator phase, the orientational hopping of EPD and ESG shapes were also analyzed within the local neighbors. At each simulation frame, the neighbor particles in the first coordination shell (for FCC crystal, it was 12) of each particle were categorized based on unique orientations and the number of each type was recorded. The normalized counts for particles with a specific unique orientation appeared to be constant and were equal to $\sim 1/N_{\Omega}$. This behavior observed in the entire trajectory. The discrete orientational jumps of neighbor particles from orientation $\Omega_m$ to $\Omega_n$, where $m, n = 1, 2, \ldots, N_{\Omega}$, were also monitored between two consecutive simulation frames. For statistical analysis, this data was collected with an optimal interval of $1\times 10^{4}$ MCS, along the MC trajectory. All possible orientational hopping were classified and counted. The normalized counts fluctuated uniformly at a value slightly less than $1/N_{\Omega}$. It meant that a single instant of hopping of the central particle, in general, did not always influence the orientations of all of its immediate neighbors. But eventually, over the entire length of an equilibrium trajectory, the discrete hopping occurred uniformly for all particles in the system. The analyses indicated that the discrete hopping of the neighbor particles occurred in manners consistent with the equipartition of all particles within the unique orientations at all times. We also implemented the transition matrix analysis in the TC system beyond the plastic phase. Despite the occurrence of two unique orientations in the BCT crystal, no discrete hopping was observed at any packing fraction in the orientationally ordered crystal, instead, all the particles remained frozen at their respective orientations with local small amplitude fluctuations.

Discrete transitions between chosen orientations for the ESG shape exhibited specific patterns and deserved further discussion. In the orientational space spanned by $\Theta_1$, $\Theta_2$ and $\Theta_3$, the ESG particles depicted six unique orientations at $\phi$ $\sim$ 0.74 as shown in the third row of Fig.\,\ref{fig:pairwise_distbn_unique_orientations_snapshots}C. The different clouds of unique orientations appeared as pairs based on their location in the orientational space of $\Theta_1$, $\Theta_2$ and $\Theta_3$. Three pairs out of six clouds of orientations were separated by $\sim$ 45$^\circ$ within a pair, but differed by $\sim$ 90$^\circ$ across the pairs. As the system melted slightly, at $\phi$ $\sim$ 0.706, 0.688, maximum number of particles had no appreciable orientational mobility, while a few particles started to toggle within the unique orientations separated by $\theta_{ij}\sim$ 45$^\circ$, shown in Fig.\,\ref{fig:discrete_mobility_hopping_matrix}B. At $\phi$ $\sim$ 0.706, the probability of hopping within the pairs was much higher than the probability across the pairs, separated by $\theta_{ij}\sim$ 90$^\circ$. The difference between two kinds of hopping, i.e.,within a pair and across the pairs, decreased at $\phi$ $\sim$ 0.688, but still showed a clear difference in the off-diagonal populations. This difference almost went to null at $\phi$ $\sim$ 0.662, showing no significant change in probabilities of hopping in the off-diagonal elements of the transition matrix (Fig.\ref{fig:discrete_mobility_hopping_matrix}B). For the ESG shape, there was a clear signature of another level of orientational coupling obeyed by all the particles in the system as indicated by our analyses.

\subsection{System size dependence of orientational behavior}
We carried out additional calculations to check the system size dependence of the unique features of the high pressure orientational phase in ESG and EPD, namely the fixed number of unique orientations with constant orientational differences and equal distributions of particles among them. Simulations with both small ($2197$ particles) and much larger ($27000$ particles) systems confirmed all attributes of the phase and ruled out any systematic variation with the size of the simulation system. For further verification of the equilibrium nature of this phase, two BCT crystalline systems at $\phi$ $\sim$ $0.707$ and $0.73$ were manually prepared: (i) one unique orientation for all the particles and (ii) two unique orientations keeping exactly equal population densities. These states were never obtained in self-assembly simulations, therefore, it could serve as different, and possibly ``non-equilibrium'' initial conditions. The \textit{NPT} runs at $P^{*}$=50, starting from these artificial states, were found to result in FCC crystals with six unique orientations for both cases, satisfying all the orientational attributes observed in the spontaneous self-assembly simulations. All the standard signatures of the equilibrium phase were verified computationally by implementing our new analyses protocol. 

\begin{figure*}
	\centering 
	\includegraphics[scale=0.9]{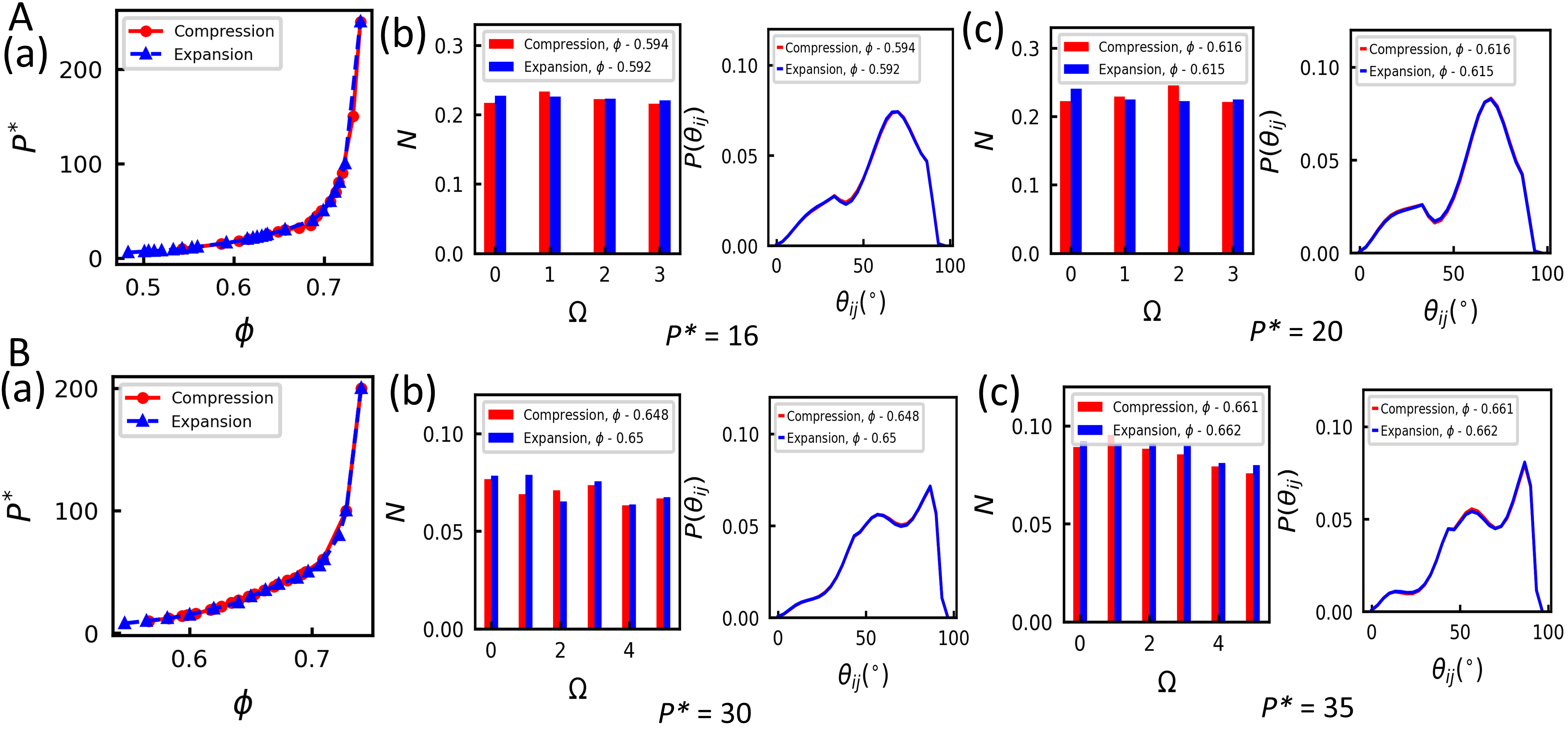}
	\caption{\textbf{Equation of states plots and the analysis of the orientational phases at two packing fractions were shown for the each of the two systems}: \textbf{(A)} EPD and \textbf{(B)} ESG. The equation of states for the system of EPD and ESG shapes are shown in panel (a) indicating no signature of hysteresis under \textit{NPT} ensemble, where the data of the solid parts are reported only. The population densities in the four and six unique orientations coordinate with each other for EPD and ESG shapes respectively, at two $P^{*}$ values each (left part of panel (b) and (c) respectively) around the transitions between the discrete rotator phase and plastic crystal phase. The distributions of all pairwise angles in the corresponding systems are displayed for both the compression and expansion runs separately at the same two $P^{*}$ values (right part of panel (b) and (c)). The orientational analysis near the transitions (panel (b) and (c)) agreed well with non-existence of hysteresis in the equation of states plots (panel (a)) for both the shapes.}
	\label{fig:eos}
\end{figure*}

\subsection{Hysteresis analysis  across the solid-solid transitions}
Analysis presented so far confirmed distinct differences between the high pressure solid region consisting of the discrete rotator phase and the freely rotating plastic phase. However, they were unable to provide information about the nature of the transition between these two phases. While in the absence of rigorous free energy calculations, it was difficult to ascertain the true nature of the transition, it was possible to look for certain signatures of the first order transition by using the methods presented here. The most obvious marker was the presence of hysteresis in thermodynamic and structural properties across the phase boundary. We performed slow compression simulations under constant pressure conditions starting from the lowest density plastic crystalline solids, with compression rates comparable with the expansion or melting rates.

Purely orientational transitions in the EPD and ESG systems did not show any hysteresis in the equations of states (Fig.\,\ref{fig:eos}A(a) and B(a)). We further compared orientational attributes of the systems between the expansion and the compression runs, near the approximate phase boundaries, ascertained from the distributions of $\theta_{ij}$ (Fig.\,\ref{fig:pairwise_distbn_unique_orientations_snapshots}). While calculating unique orientations in the compression and expansion runs, fixed reference quaternions, corresponding to the special orientations $\mathbb{Q}$, were taken from the high pressure systems. Data for two pressure values are presented in Fig.\,\ref{fig:eos}A(b)-(c) and B(b)-(c). In both systems, and at both the pressure values, the same number of unique orientations with equal populations were found, within statistical error (left subfigures of Fig.\,\ref{fig:eos}A(b)-(c) and B(b)-(c)). The distributions of all pairwise angles $\theta_{ij}$ were also exactly same (right subfigures of Fig.\,\ref{fig:eos}A(b)-(c) and B(b)-(c)). These results indicated no signature of hysteresis across the the solid-solid transitions.

\begin{figure*}
	\centering
	\includegraphics[scale=0.23]{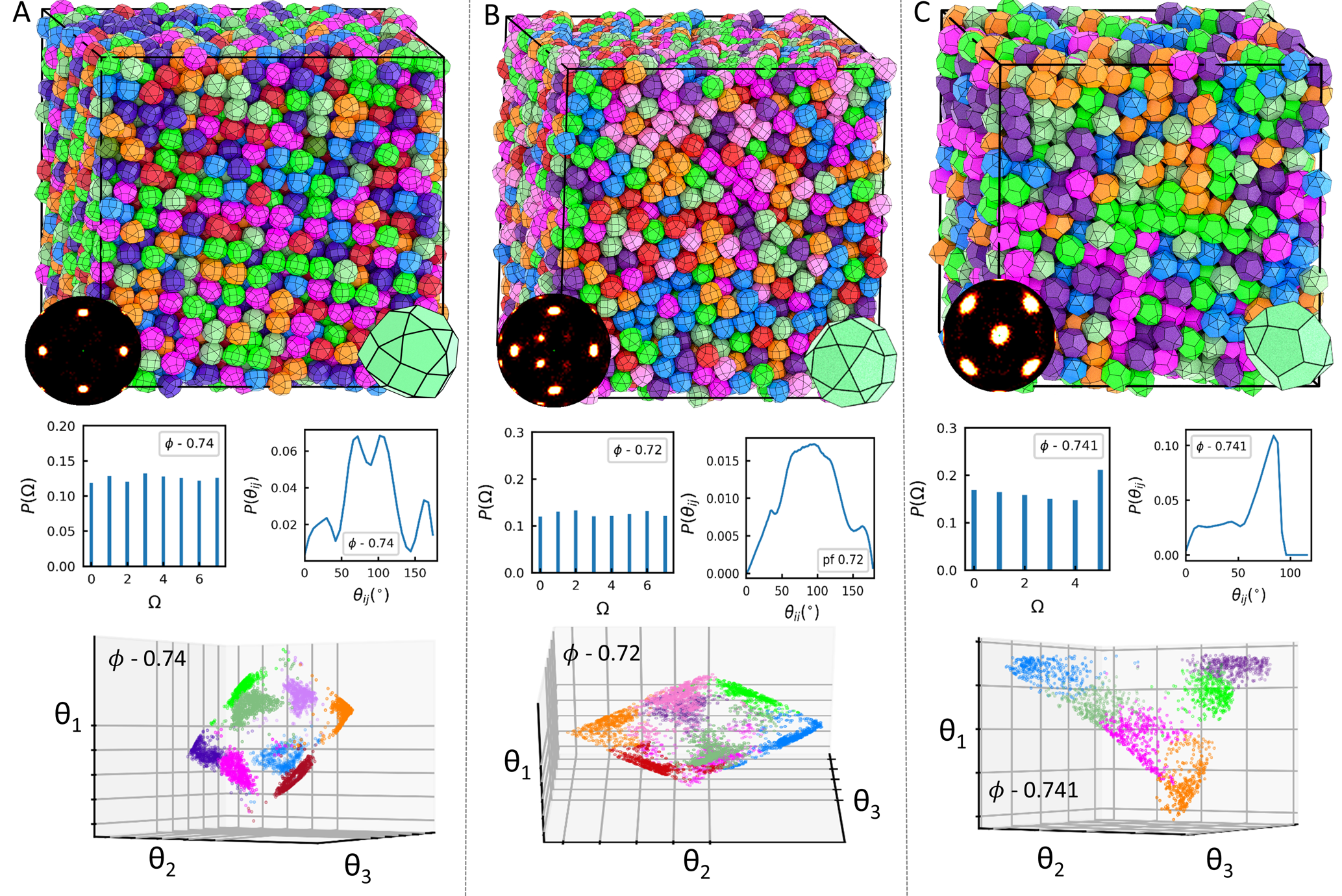}
	\caption{\textbf{Orientational disordered phase in the self-assembled crystals are shown for three polyhedral shapes:} \textbf{(A)} Elongated Pentagonal Gyrocupolarotunda (EPG), \textbf{(B)} Elongated Pentagonal Orthocupolarotunda (EPO), and \textbf{(C)} Parabiaugmented Dodecahedron (PD). The simulation snapshots are shown in multiple colors at the densest states (particle shape and bond-orientational order diagram of the crystal are shown as inset), where the number of unique orientations appear to be eight, eight and six for EPG, EPO and PD respectively as indicated by the clouds in the three-dimensional distributions. Equal partitioning within the unique orientations and fixed orientational differences are maintained for the three shapes within the statistical noise. All the attributes remain preserved for these three shapes leading to the existence of qualitatively identical features of orientational disorder as observed for EPD and ESG shapes in Fig.\,\ref{fig:pairwise_distbn_unique_orientations_snapshots}.} 
	\label{fig:shape_space}
\end{figure*}

\subsection{Orientational attributes in the disordered crystals of other shapes}
The orientationally disordered phase could be characterized in the unit cell of the crystals by the presence of specific number of unique orientations with fixed differences and equal partitioning of particles among them. Both the EPD and the ESG shapes fit this general qualitative description. The actual number of orientations was a quantitative detail that was the property of the shape. We analyzed the orientational phase behaviors of crystals formed by hard convex polyhedra as reported by Damasceno \textsl{et\,al.}\cite{Damasceno2012c}. Three more shapes, EPG, EPO and PD, with similar orientational features were identified. The results are presented in Fig.\,\ref{fig:shape_space}. Among the three shapes, EPG and EPO shapes were found to form FCC crystalline ordering and the PD shape self-assembled into body centered cubic (BCC) crystal, confirming the assignments reported in the literature \cite{Damasceno2012c}. All the shapes exhibited freely rotating plastic crystal phases at low pressure regions and maintained the same positional ordering across the phase diagrams, giving rise to purely orientational solid-solid transitions. We detected unique orientations with angular cut-offs of 50$^{\circ}$, 45$^{\circ}$, 45$^{\circ}$ for EPG, EPO and PD shapes, respectively. All three additional shapes contained small number of special orientations which were eight, eight and six for the EPG, EPO and PD shapes, respectively (Fig.\,\ref{fig:shape_space}). The uniquely oriented particles were found to be randomly distributed across the systems without any clustering, as found in the EPD and ESG systems. Furthermore, the equal partitioning among the special orientations, was found to be present as well (Fig.\,\ref{fig:shape_space}). In summary, the same orientational features were preserved in the crystalline systems of three convex polyhedra under hard-core interaction.

\begin{figure*}
	\centering
	\includegraphics[scale=0.8]{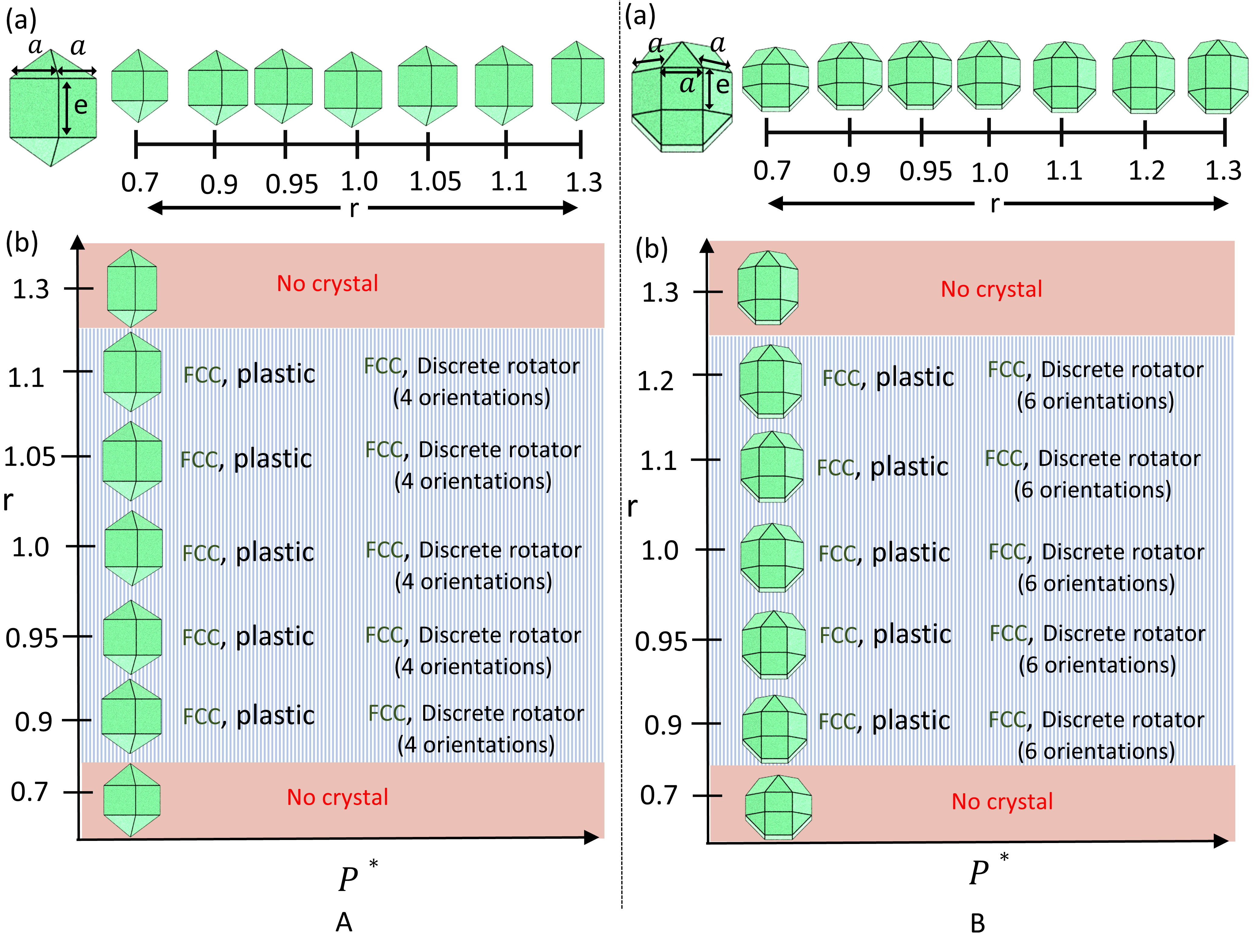}
	\caption{\textbf{The degree of elongation and translational/orientational order in the phase diagram were shown:} \textbf{(A)} EPD and \textbf{(B)} ESG shape. The elongation/contraction parameter ``$r$'' was defined as the ratio of ``$e$'' and ``$a$'' as shown in the panel (a). The degree of elongation for the EPD and ESG shapes were varied by changing the value of ``$r$'' within 0.7 to 1.3. $r = 1$ meant the regular shape with all edge lengths equal; $r > 1$ and $r < 1$ indicated the ``elongated'' and ``contracted'' shape respectively. The translational and orientational order in the lower and higher pressure regions, exhibited by the elongated/contracted EPD and ESG shapes, are shown in the shape space with a variation of $r$ (panel (b)). It suggests the robustness of the phase which is independent of the degree of elongation for the polyhedral shapes.} 
	\label{fig:models}
\end{figure*}

We expanded the search by performing systematic modifications to the EPD and ESG shapes, followed by simulations and analyses by the same protocols. The shapes were both ``elongated'' and ``contracted'' by geometric operations depicted in Figs.\,\ref{fig:models}A(a) and B(a). This gave us two families of convex polyhedral shapes around the EPD and the ESG. A shape in this family is characterized by a parameter $r$, defined in Figs.\,\ref{fig:models}A(a) and B(a). We considered altogether twelve shapes, six in each case, covering the expanded and contracted regions of the shape space. All shapes in both EPD and ESG families, except for the four at both extremes, exhibited identical orientational behavior with the same attributes as the native polyhedra (Figs.\,\ref{fig:models}A(b) and B(b)). The shapes at the extreme ends of each family ($r$ = 1.3 and 0.7) did not crystallize even after $\sim$ 150 million MC runs. These results clearly indicate the robustness of the phase behavior under systematic modifications of particle shapes in the context of crystals of hard convex polyhedra.

\subsection{Particle symmetry and the existence of the disordered phase}\label{symm_relation}
Additional data underscored the fact that the disordered orientational phase with discrete mobility is quite common in purely entropic model systems of crystals and search for similar behavior could be an exciting new direction in nanoscience. From the standpoint of materials design, it is natural to look for common features in all the occurrences, especially, in terms of attributes of particle shapes. All the shapes used in this study self-assembled into cubic crystals (FCC or BCC). From the perspective of the particle shapes, we observed a commonality related to the point group symmetries of the anisotropic bodies. The role of point groups of the anisotropic particles and local environments was reported earlier \cite{Shen2019, Lee2023}. In these studies, the existence of discrete rotator phase or continuous rotator phase were classified in terms of the commonality of certain rotational operations in the point groups of the particles and local environments. In the three dimensions, for the polyhedral shapes used in our study, the point groups could be characterized as ones with relatively ``low symmetries'', compared to the high symmetry octahedral or icosahedral points groups. On top of that, all the particle point groups belonged to the category of non-crystallographic point groups (Table \ref{table:symmetry_table}). Though, we were unable to find any other straightforward relationships based on these observations, we hypothesize that if a polyhedral shape with non-crystallographic point group symmetry self-assembles into any cubic crystal, there is a strong possibility for the finite set of unique orientations occurring in the disordered structure due to the ``symmetry mismatch'' between the particle and crystal. In this phase, the unique orientations were arranged in such way, some specific symmetries of the crystal structure could be broken following a particular manner giving rise to a set of discrete orientations. This hypothesis was supported by the data in the shape family based on the EPD and ESG by varying the geometry of the shapes in a controlled way. Modified shapes, irrespective of the elongation or contraction, i.e.\,the value of the parameter $r$, preserved the point group symmetries of the native shapes. This fact underscored the importance of particle symmetry, above other geometrical attributes, as a predictive marker for the orientational phase behavior of the crystal structure formed by the particle in question. We like to point out that the evidence presented here is empirical and does not rule out other situations where this phase could materialize.

\begin{table*}[t]
	\centering
	{\renewcommand{\arraystretch}{1.0}%
		\begin{tabular}[t]{|>{\centering\arraybackslash}m{4.5cm}|>{\centering\arraybackslash}m{4cm}|>{\centering\arraybackslash}m{3cm}|>{\centering\arraybackslash}m{4cm}|}
			\hline
			Shape name (Abbreviation) & Point group symmetry of particle & Number of unique orientations (Higher $P^{*}$) & Existence of the discrete rotator phase/Plastic phase \\
			\hline
			Elongated Pentagonal Dipyramid (EPD) & $D_{5h}$ (20) & 4 & Yes/Yes \\
			\hline
			Elongated Square Gyrobicupola (ESG) & $D_{4d}$ (16) & 6 & Yes/Yes \\
			\hline
			Elongated Pentagonal Orthocupolarotunda (EPO) & $C_{5v}$ (10) & 8 & Yes/Yes \\
			\hline
			Elongated Pentagonal Gyrocupolarotunda (EPG) & $C_{5v}$ (10) & 8 & Yes/Yes \\
			\hline
			Parabiaugmented Dodecahedron (PD) & $D_{5d}$ (20) & 6 & Yes/Yes \\
			\hline
			
		\end{tabular}
		\caption{The polyhedral shapes used in the study are included with the point group symmetry and total number of operations. The existence of the discrete rotator phase and plastic crystal along with the number of unique orientations found in high pressure solid regions are also displayed.}
		\label{table:symmetry_table}
	}
\end{table*}

\section{Discussion}\label{discussion}
The primary technical advancement reported in the paper was the algorithmic procedures to detect unique orientations in a robust and exhaustive fashion. A version of the method, which was amenable to visual inspection of the many-body behavior of the particle orientations by projecting them on a three dimensional space of orientational differences from three reference values, was helpful not only for detecting the number of orientations, but the fluctuations around the discrete values. The inference from quantitative estimations could always be verified by computer graphics of simulation snapshots where particles were colored by their unique orientation. The complete statistics of orientations enabled us to rigorously estimate the discrete hopping behavior in terms of particle and ensemble averaged transition probabilities. Together, the procedure outlined in this paper, which was developed based on the idea presented in the work of Karas \textsl{et al.} \cite{Karas2019}, provide a powerful approach to study orientational behavior of any many-body system consists of rigid bodies.

Before we move on to further discussions of important findings and implications of the paper, we present a brief comparison with other methodological approaches to handle rotational motions of symmetric object in the context of equilibrium phase behavior of many-body systems. The point group symmetry of the rigid body necessitates additional considerations to be dealt with while representing particle orientations. This is especially true when the final objective is to formulate a notion of comparison involving rotational states of two particles. In our approach, we handled the problem computationally by taking the minimum of all possible quaternion angles between a pair of particles, after exercising the rotational operations of the point group in question. The rest of the technicalities in our method follows quite directly from this notion of orientational difference based on the quaternion representation of the continuous rotation group. While the angle was well defined, it did not have the complete information of the rotational states of the two particles in question, because the axis information was not explicitly used. In order to handle this while identifying the unique orientations, we used three reference quaternions, which took care of the problem of uniqueness, or, only focused on the limit $\theta_{ij}\sim 0$, which essentially produced the same end result. Several recent formulations used a irreducible set of linearly independent vectors for each rigid body to represent its orientation. In this setting, in order to handle the restrictions posed by the particle point group in a rigorous fashion, it was necessary to go to a tensor representation \cite{Akbari2015, Nissinen2016}. In these frameworks, the orientation of a symmetric object with specific point group was expressed as a tensor obtained from the set of vectors mentioned above. Detailed prescriptions for the construction of respective tensors for families of points groups were presented in these papers \cite{Akbari2015, Nissinen2016}. In the SOC formulation, Haji-Akbari \textsl{et al.}\cite{Akbari2015} showed that the orientational difference of two symmetric bodies could be uniquely determined in terms of Frobenius norm of the difference between their respective tensorial coordinates. Hence, the tensor based SOC approach could be used to estimate the extent of orientational order and disorder. These methods are mathematically equivalent to ours and are alternative means to obtain the same information presented in this paper.

In the light of our newly introduced analysis, it could be inferred that the discretely mobile nature of particle orientations was a characteristics of a proper equilibrium thermodynamic phase and not an arrested state, despite the limited mobility at the higher packing fractions. This is the three dimensional analogue of the phase identified by Shen \textsl{et al.}\,in two dimensions \cite{Shen2019}. We speculate that the recently reported colloidal clathrates \cite{Lee2023} will show similar behavior if subjected to the analysis protocol presented here. The defining features of this phase are the following: 1) presence of a finite set of discrete orientations, 2) equal populations within the unique orientations and 3) constant pairwise orientational differences. These features were reflected in the distribution of all pairwise angles in the system and were always conserved along the entire packing packing fraction range of this phase. The discrete mobility was a result of them and manifested in the low density solids. At high pressure, when the mobility was absent, the particles did not freeze in random orientations but in exactly those values where the mobility existed at lower packing fractions. The entire range of behavior was important to identify the phase making it qualitatively distinct from the freely rotating plastic phase or some other unspecified form of disorder which varied continuously with pressure without any discernible conserved behavior \cite{Ni2012}. 

It is important to note that the mere appearance of few distinct orientations is not enough to ensure that the system under consideration belongs to the discretely mobile phase. It can, in principle, occur in out-of-equilibrium systems and may not follow a well-defined behavior across the pressure range, for example, in the orientational glass phase of mixed crystals, or the recently observed nanocrystal superlattice with frozen orientations \cite{Abbas2022}. The equilibrium nature of the system can only be verified if the complete statistics of the orientational behavior and direct evidence of mobility are collected. These arguments underscore the importance of collecting proper statistics, which is not trivial to do, unless an automatic procedure to detect unique orientations is employed, thereby justifying the methodological advances reported in this paper. Even for equilibrium systems, discrete orientations do not imply discrete mobility, as illustrated by the orientationally ordered BCT phase of the TC shape with two distinct values. Three defining features, mentioned above, seemed to be rigorous signature of this phase and discrete rotational mobility followed from these. The features were always conserved even when the system lost rotational motion in the high pressure regions of the phase diagram.

We introduced a clear notion of orientational order and disorder in crystalline system, which takes into account of the positional structure of crystals. In the light of notion and the detailed analysis presented in the paper, the true nature of disorder of the discrete plastic phase became clearer. The global disorder and random arrangements at the level of the unit cell level is an important feature of this phase, and makes it distinct from other forms of disorder.

The mismatch in symmetry has been pointed out in the polygon case \cite{Shen2019}, but it was not clear if the same argument would hold for polyhedra, although recent work on colloidal clathrate seemed to indicate that \cite{Lee2023} certain generalizations could be possible. We investigated the relationship between the properties of the particle point group and the existence of the discrete rotator phase. The empirical findings mentioned in Section \ref{symm_relation} can also be cast as a symmetry mismatch argument, which is different from the one reported earlier \cite{Shen2019}. We observed the difference between the rotational symmetry elements of the particle point group and crystallographic point group. It is plausible that this mismatch can encompass the same effect in terms of local environment, but the generality of that statement is not immediately obvious. One intriguing feature of the particle point groups that gave the discrete rotator phase was the fact they were all non-crystallographic in nature. From this limited observations, it is not clear whether this is a general universal attribute, and more investigation is needed. Nonetheless, the empirical relationship between the non-crystallographic point group symmetry of the particle and the existence of the discrete rotator phase suggested a straightforward design principle for realizing this phase in real colloidal crystals or NCSLs. Several aspects of the phase can be exploited for materials applications, for example, a subset of the special orientations could be coupled to some external stimulus to achieve specific applications. The predictive hypothesis presented here, especially the role of symmetry, could also be used as a guiding principle for fundamental understanding of the phase behavior of hard polyhedral systems and could open up new directions of computational and theoretical research.

It is customary to invoke the obvious notion of face-to-face contacts in understanding phase behavior of hard polyhedral particles \cite{Teich2019,Vo2022,Harper2019,Greg2015}. We could not find a direct connection between the nature of unique orientations and the packing behavior. For example, it did not seem that contact maximization was the main controlling factor behind the specific particle orientations. Visual inspections of unit cells of two systems (Figs.\,\ref{fig:unit_cell}B and C) revealed that the ESG shape was much better packed than the EPD in terms of face-to-face contact maximization. This behavior could be understood in the light of the symmetry mismatch argument. Due to the presence of the five-fold rotational symmetry axis ($C_5$) in EPD (point group: $D_{5h}$), the mismatch was maximal, as the crystallographic point group $O_h$ did not posses this. Whereas, ESG shape did not exhibit this behavior, because all rotational elements of the particle point group ($D_{4d}$) existed in the $O_h$ point group as well.

Based on the data provided here we can speculate on the nature of solid-solid transition, although a full account of this issue can only be completed with rigorous free energy calculations, which is not a straightforward business in disordered crystals as we discuss below. We found no system size dependence, among 2197, 4096 and 27000 particle systems, in the spatial arrangements of the unique orientations across the systems. Other aspects of the conserved attributes and orientational disorder were also independent of system size. We estimated the thermodynamic compressibility with external pressure and found no signature of divergence and system size dependence of the transition for ESG shape. These observations ruled out a straightforward occurrence of a second order continuous transition. The qualitative nature of the orientational behaviors suggested the possibility of the transition being first order. However, we found no signature of hysteresis in any of the markers of the phases, as well as in the generic feature, namely the equation of state. However, the lack of the hysteresis does not necessarily indicate the absence of a first order transition. It is possible that, for a sufficiently low free energy barrier, the hysteresis will be either small or completely non-existent. Similar qualitative observation could be found in the transition between the freely rotating plastic phase and the ordered phase, detailed analysis of which in the Rhombic Dodecahedron and Rhombicuboctahedron shape has been reported before \cite{Karas2019}. Prominent signatures of hysteresis loop in the equation of state was only found in cases where a positional transition was taking place. For example, in the TC shape between the FCC and the BCT structures \cite{Karas2019}. It is conceivable that the free energy barrier involving a change in positional structure is much larger than that of a purely orientational change occurring on the same crystal lattice. 

For quantitative assessments of relative stability of phases and nature of transitions, proper free energy calculations are needed. Application of the standard Frenkel-Ladd method \cite{Frenkel1984, Vega2007} in the presence of complex orientational disorder is not straightforward. Such extension has been carried out in determining the comparative stability between plastic phases with different crystal structures with an orientationally ordered reference state \cite{Marechal2008}. In another study, two distinct solid phases with somewhat comparable orientations were studied with a similar approach \cite{Amir2011}. A reversible path for the thermodynamic transformation between the ordered to the freely rotating plastic phases could be constructed in the Frenkel-Ladd formulation, because of the ``complete'' disorder of the target phase. However, for the kind of site specific disorder with several distinct orientations present in the discrete rotator phase, such construction is not obvious. Even if one could realize this, the large degeneracy originating from the disorder poses a separate problem. The combinatorial complexity is very large in determining the number of microstates that are consistent with the conserved quantities, despite their weights being equal. This fact prevents one from using the approach proposed by Pauling to estimate ground state entropy of the ice crystal with disordered orientations of the water molecules\cite{Pauling1935}. One possible way forward is to figure out the degeneracy by mapping the problem to another model and then use an extension of the Frenkel-Ladd method with properly constructed thermodynamic path. Both of which require significant methodological advances and will be explored in future work.

The distinct nature of this phase can be understood in terms of broken symmetry of rotational motion of a single particle in the crystal structure. The following statements can be made about each realization of the system. In the plastic crystal phase, rotational motion is continuous and in the local coordinate system of a particle, the symmetry is not broken. On the contrary, in the discrete rotator phase, the symmetry is broken in a very specific manner characterized by the number of unique orientations in the specific translational order depicted by the crystal structure. This symmetry breaking is different from the usual broken rotational symmetry of the entire crystal, and is associated with the conserved quantity mentioned above. This line of argument provided additional justification to categorize different behavior as a separate thermodynamic phase in the absence of further quantitative details.

It was quite apparent that the discrete rotator phase was a result of nontrivial correlation in the many-body system. While we could shed any light on the true nature of the correlation, its manifestation were well defined in terms of the statistics of the rotational motions of the particles. Despite equilibrium fluctuations, particle motions maintained the conserved quantities discussed above. The signatures of correlations were always present even when the rotational mobility of the particles were absent as it was the case in the high pressure regions of the phase diagram. It is important to note, that discretely mobile behavior in low density solid does not necessarily guarantee that in the frozen state at higher packing fraction the particles will adopt the same exact orientations, maintaining all the conserved features across the system. This is only possible if the mobile low density solid and the frozen state at the higher densities were essentially the same thermodynamic state with the same underlying inter-particle correlations. This observation made the disorder non-trivial, and warranted a separate thermodynamic description from the freely rotating plastic phase, in other words, indicated an existence of solid-solid transition which was purely orientational in nature. Based on these arguments, the discrete rotator phase could also be interpreted as a case of ``correlated disorder'', in line with the well studied phenomenon observed in varieties of strongly correlated multi-component atomic crystals \cite{Chaney2021, Meekel2021, Simonov2020}. The appearance of correlated disorder is associated with existence of a ``local rule'' which the system always obeys while manifesting the disordered configurations. The earliest example of this is the Bernal-Fowler \cite{Bernal1933} rule in ice crystal which underpins the orientational disorder of water molecules and was the basis of entropy estimation by Pauling mentioned earlier\cite{Pauling1935}. This local rule is the source of correlation which essentially gives rise to the novel disorder because of entropy maximization. It is intriguing to speculate whether such rules exists in the orientational space of the polyhedra for the discrete rotator phase. This could explain the emergence of the conserved quantities reported in the paper. A detailed study along this line will be presented in a future publication.

\section{Conclusions}
Our results suggest that purely entropy driven classical systems can have a rich phase behavior with non-trivial disordered ground states at certain parts of the phase diagram. The disordered phase was characterized by discrete rotational motions of the particles, which made it stand out from the usual form disorder with continuous, fully random orientations. This discrete behavior in a classical system, with simultaneous occurrence of certain conserved quantities, and effectively infinite ranged orientational correlation were quite striking from a fundamental standpoint. Recent interests in understanding orientational phase behavior of colloidal crystals or NCSLs, coupled with wealth of information on the self-assembly of anisotropic building blocks, could provide the test bed for direct experimental validations of the predictions discussed in this paper. Ideas presented here could also be taken up as inspirations for synthesis of anisotropic nanoparticles and designing new protocols for making target crystalline structures. New evidence of non-trivial behavior in hard particle classical systems further underscores the versatility of entropy, without energy, as an organizing principle \cite{Onsager1949,Frenkel1999}, and particle shape as a design principle \cite{Onsager1949,Glotzer2007}.

\begin{acknowledgements}
	Authors thank Prof.\,Sharon Glotzer and members of the Glotzer group for insightful discussions. SK acknowledges financial support from DST-INSPIRE Fellowship (IVR No.\,201800024677). AD thanks DST-SERB Ramanujan Fellowship (SB/S2/RJN-129/2016) and IACS start-up grant. KC thanks IACS for financial support. Computational resources were provided by the IACS HPC cluster and partial use of equipment procured under SERB CORE Grant No.\,CRG/2019/006418.	
\end{acknowledgements}

\section*{AUTHOR DECLARATIONS}
\subsection*{Conflict of Interest}
The authors have no conflicts to disclose.

\subsection*{Author Contributions}
SK, KC and AD designed the research. SK and KC performed all the simulations, SK, KC and AD analyzed the data. SK and AD wrote the paper.

\bibliography{correlated_disorder_v1}

\end{document}